\documentclass[a4paper,12pt]{article}
\usepackage{latexsym}
\usepackage{amsfonts} 
\usepackage[mathscr]{euscript}
\usepackage{amsmath,mathrsfs,bm,amssymb,color}

\author{
Tadahiro Miyao\footnotemark[1]
  and  Herbert Spohn\footnotemark[2]\\ 
{\it Zentrum Mathematik,}
{\it Technische Universit\"at M\"unchen,}\\ 
{\it  D-85747 Garching, Germany}\\
e-mail:
\footnotemark[1] miyao@ma.tum.de,
\footnotemark[2] spohn@ma.tum.de }

\title{\textsf{Spectral Analysis of the Semi-relativistic Pauli-Fierz Hamiltonian}} 
\date{\empty}

\newcommand{\one}{{\mathchoice {\rm 1\mskip-4mu l} {\rm 1\mskip-4mu l} 
{\rm 1\mskip-4.5mu l} {\rm 1\mskip-5mu l}}} 
\newcommand{\h}{\mathfrak{h}} 
 
\newcommand{\ex}{\mathrm{e}} 
\newcommand{\D}{\mathrm{dom}} 
 
\newcommand{\im}{\mathrm{i}} 
\newcommand{\Fock}{\mathfrak{F}} 
\newcommand{\Ffin}{\mathfrak{F}_{\mathrm{fin}}} 
\newcommand{\BGamma}{\check{\Gamma}} 
\newcommand{\dG}{\mathrm{d}\Gamma}

\newcommand{\ran}{\mathrm{ran}} 
 
\newcommand{\la}{\langle} 
\newcommand{\ra}{\rangle} 
 
\newcommand{\wlim}{\mbox{$\mathrm{w}$-$\displaystyle\lim_{n\to\infty}$}}

\newcommand{\BbbR}{\mathbb{R}} 
\newcommand{\BbbN}{\mathbb{N}}

\newcommand{\BbbC}{\mathbb{C}} 
\newcommand{\vepsilon}{\varepsilon} 
\newcommand{\vphi}{\varphi} 
 
\newcommand{\Pf}{P_{\mathrm{f}}} 
\newcommand{\Nf}{N_{\mathrm{f}}} 
\newcommand{\Hf}{H_{\mathrm{f}}}

\newcommand{\mph}{m_{\mathrm{ph}}} 
\newcommand{\Dirac}{\tilde{D}^{\otimes}(P)} 
\newcommand{\A}{A(0) }
\newcommand{\B}{B(0)} 
\newcommand{\C}{\mathrm{const}\, } 
\newcommand{\J}{\mathscr{J}} 
\newcommand{\Od}{\mathcal{O}} 
\newcommand{\s}{S_{k, \lambda}} 
\newcommand{\EU}{\mathcal{E}_-} 
\newcommand{\EO}{\mathcal{E}_+} 
 
\def\Sumoplus{\sideset{}{^{\oplus}_{n\ge 0}}\sum}

\begin{document}

\newtheorem{define}{Definition}[section] 
\newtheorem{Thm}[define]{Theorem} 
\newtheorem{Prop}[define]{Proposition} 
\newtheorem{lemm}[define]{Lemma} 
\newtheorem{rem}[define]{Remark} 
\newtheorem{assum}{Condition} 
\newtheorem{example}{Example} 
\newtheorem{coro}[define]{Corollary}

\maketitle 

\begin{abstract}
We consider a charged particle, spin $\frac{1}{2}$, with relativistic kinetic energy and minimally coupled to the quantized Maxwell field. Since 
the total momentum  is conserved, the Hamiltonian admits a fiber decomposition as $H(P)$, $P\in \BbbR^3$. We study  the spectrum of $H(P)$.
 In particular we  prove that, for non-zero photon mass, the ground state is exactly two-fold degenerate and separated  by a gap, uniformly in $P$, from 
the rest of the spectrum.
\end{abstract}

 \section{Introduction and Main Results }
 Let us consider a  classical point charge, charge $e$, mass $M$, position $q$, velocity $\dot{q}$, coupled to the Maxwell field with electric field $E$
 and magnetic field $B$. The coupling to the field is through a rigid charge distribution $\vphi: \BbbR^3\to \BbbR_+$ normalized as $\int\mathrm{d}x\, \vphi(x)=1$. Then the equations of motion for the coupled system read, in units where $c=1$,
 \begin{align}
 &\frac{\partial}{\partial t}B=-\nabla \wedge E,\ \ \ \  \frac{\partial}{\partial t} B=\nabla\wedge E-e \vphi(\cdot-q)\dot{q},\nonumber\\
 &\nabla\cdot E=e\vphi(\cdot-q),\ \ \ \  \nabla\cdot B=0,\nonumber\\
 &\frac{\mathrm{d}}{\mathrm{d}t}\big(M (1-\dot{q}^2)^{-1/2}\dot{q}\big)=e\big(E*\vphi(q)+\dot{q}\wedge (B* \vphi)(q)\big)\label{Maxwell3}
 \end{align}
 with $*$ denoting convolution. The uncoupled system, $e=0$, is Lorentz invariant. But the choice of the rigid charge distribution singles out  a specific reference frame and hence makes the model semi-relativistic, only.
 
 The canonical quantization of (\ref{Maxwell3}) results in a quantum evolution governed by the semi-relativistic Pauli-Fierz hamiltonian.
 Our goal is to study spectral properties of this operator. While the nonrelativistic counterpart has been investigated in considerable detail, no spectral results
  seem to  be available for the semi-relativistic case.

  The quantization procedure for (\ref{Maxwell3}) is described , e.g., in \cite{Spohn2}. One writes  (\ref{Maxwell3}) in  Lagrangian form and Legendre transforms to Hamiltonian
  structure in using the Coulomb gauge. Since our prime example will be an electron (charge $-e$), we want to include spin $\frac{1}{2}$. As for the 
  nonrelativistic hamiltonian this  amounts to replacing $(p+eA)^2$ by $\big(\sigma\cdot (p+eA)\big)^2$ with $\sigma$ the $3$-vector of Pauli spin matrices.
  As a result one obtains the semi-relativistic Pauli-Fierz hamiltonian, which is given by
  \begin{align}
H=\gamma \sqrt{\big(\sigma\cdot(-\im \nabla_x+eA(x))\big)^2+M^2}+H_{\mathrm{f}}.\label{IntMainFull}
\end{align}
We introduced here by hand the factor $\gamma,\ 0<\gamma\le 1$.  Physically, as discussed above, one has $\gamma =1$. Our units are 
chosen such that $\hbar =1$.
 Without restriction of generality we set $e\ge 0$.
 $H$ acts in $L^2(\BbbR^3; \BbbC^2)\otimes \Fock$, where $\Fock$  is  the photon Fock space
\[
 \Fock=\Sumoplus L^2(\BbbR^3\times\{1,2\})^{\otimes_{\mathrm{s}}n}.
\]
 $A(x)$ is the quantized  vector potential defined through
\[
 A(x)=\sum_{\lambda=1,2}\int_{|k|\le
 \Lambda}\frac{\mathrm{d}k}{\sqrt{2(2\pi)^3\omega(k)}}\vepsilon(k, \lambda)\big( \ex^{\im
 k\cdot x }a(k,\lambda)+\ex^{-\im k\cdot x}a(k, \lambda)^*\big),
\]
where $\vepsilon(k, \lambda), \lambda=1,2$, is the pair of polarization vectors.  $k/|k|, \vepsilon(k, 1), \vepsilon(k, 2)$ are  a dreibein depending  
measurably on $k$. 
For convenience we use the sharp ultraviolet cutoff $\Lambda$   which corresponds to setting $\hat{ \vphi}(k)=(2\pi)^{-3/2}$ for $ |k|\le \Lambda$
and $\hat{\vphi}(k)=0$ otherwise,   $\hat{}$ denoting  Fourier transform.
Our results are equally valid for a smooth cutoff.
$a(k, \lambda), a(k, \lambda)^*$ are the annihilation
and creation operators which satisfy the standard commutation relations
\[
 [a(k, \lambda), a(k', \lambda')^*]=\delta_{\lambda\lambda'}\delta(k-k'),\
 [a(k, \lambda),a(k', \lambda')]=0=[a(k, \lambda)^*,a(k', \lambda')^*].
\]
$H_{\mathrm{f}}$ is   the field energy,
\begin{align}
 H_{\mathrm{f}}&=\sum_{\lambda=1,2}\int_{\BbbR^3}\mathrm{d}k\,
 \omega(k)a(k, \lambda)^*a(k, \lambda). \label{HfDef}
\end{align} 
For the Maxwell field the dispersion relation is 
\[
\omega(k)=|k|.
\]
Mathmatically it is convenient to introduce the photon mass  $\mph$ through the choice
\[
\omega(k)=\sqrt{k^2+\mph^2}.
\]
Readers will  find more precise  definitions of $A(x)$ and $\Hf$ in the Appendix \ref{SecondQ}.

\begin{rem}
{\rm
For a fixed configuration of the vector potential the classical hamiltonian function is 
\[
H_{\mathrm{cl}}(p, q)=\sqrt{(p-eA(q))^2+M^2}.
\]
We picked here the ``naive" quantization $p\leadsto -\im \nabla_x$, $q\leadsto x$, which is fairly common in the physics
community \cite{DL}. Alternatives would be either Weyl or magnetic Weyl quantization \cite{MP}. $\diamondsuit$
}
\end{rem}

By translation invariance the total momentum, i.e., the sum of the momentum of the charge and the  field momentum, is conserved.  The generator of translations is the total momentum operator
$P_{\mathrm{tot}}=-\im \nabla_x+\Pf$ with
\[
 \Pf=\sum_{\lambda=1,2}\int_{\BbbR^3}\mathrm{d}k\,
ka(k,\lambda)^*a(k,\lambda).
\]
It  strongly commutes with  the hamiltonian $H$,
namely, $\exp[-\im a\cdot P_{\mathrm{tot}}]\exp[-\im t H]=\exp[-\im t
H]\exp[-\im a \cdot P_{\mathrm{tot}}]$ for all $a\in \BbbR^3$ and $t\in
\BbbR$.  Therefore $H$ admits the direct integral decomposition
\begin{align} 
\mathcal{U}H\mathcal{U}^*&=\int^{\oplus}_{\BbbR^3} H(P)\, \mathrm{d}P, \label{IntMainFiberNR}\\
H(P)&=\gamma \sqrt{(P-\Pf+e\A)^2+\sigma\cdot \B+M^2}+\Hf
\end{align}
acting in $\BbbC^2\otimes \Fock$,  $B(0)=\nabla \wedge A(0)$. The unitary $\mathcal{U}$ is defined by $\mathcal{U}=\mathcal{F}_x \exp[\im x\cdot \Pf]$ where 
$\mathcal{F}_x$ is the Fourier transformation with respect to $x$. We will provide a   mathematically rigorous definition of $H$ and $H(P)$ in section \ref{SAGS}. 
Our interest is mostly in the low lying spectrum of $H(P)$ in dependence of $P$.
  To get started we have to ensure the self-adjointness of $H$ and of $H(P)$,  see section \ref{SAGS} for details.
\begin{Prop}\label{SAA} 
For any $0<\gamma \le 1, \Lambda<\infty$ and $0\le \mph$, there exists $e_*>0$ such that, for all $e<e_*$, $H$ is self-adjoint on 
 $\D(|-\im \nabla_x|)\cap \D(\Hf)$. Moreover $H$ is essentially 
 self-adjoint on any core of the free Hamiltonian $H_0=\gamma \sqrt{-\Delta_x+M^2}+\Hf$. 
\end{Prop}

\begin{Prop}\label{SAB}Choose $\gamma,  \Lambda, \mph$  arbitrarily as Proposition \ref{SAA}.  Let $e_*$ be given by Proposition 
 \ref{SAA}. Then, for all $e<e_*$ and $P\in \BbbR^3$, $H(P)$ is self-adjoint on $\D(|\Pf|) 
 \cap \D(\Hf)$. Moreover $H(P)$ is essentially self-adjoint on any core 
 of the operator $H_0(P)=\gamma \sqrt{(P-\Pf)^2+M^2}+\Hf$. 
\end{Prop}  
\begin{rem}
{\rm 
There are further parts of our proof which will require small $e_*$. Therefore we did not attempt to optimize $e_*$ in every step. $\diamondsuit$
}
\end{rem}

The spectral analysis of the nonrelativistic Pauli-Fierz hamiltonian was initiated  by J. Fr\"ohlich  in his  Ph.D. thesis  \cite{JFroehlich1}.
 Our first main result is the extension of his methods  to the semi-relativistic case. 
While the  result could be   anticipated from \cite{JFroehlich1, Moller2, Spohn1}, 
the actual proof is suprisingly technical, since the minimal coupling is under the square root,
see  section  \ref{ProofSpectrum}.  
\begin{Thm}\label{Spectrum}Set $ \Lambda, \gamma, \mph$ arbitrarily as in Proposition \ref{SAB}.
Choose $e$ as $e<e_*$. Let 
 \begin{align*}
\Delta(P)=\inf_{k\in \BbbR^3} \big( E(P-k)+\omega(k)-E(P)\big)
\end{align*}
where
\begin{align*}
E(P)=\inf \mathrm{spec}(H(P)),\ \ \ \Sigma(P)=\inf \mathrm{ess. spec}(H(P)).
\end{align*} 
 Then one has
\begin{align*} 
\Sigma(P)-E(P)= \Delta(P)
\end{align*} 
for all $P\in\BbbR^3$.
\end{Thm} 

A problem of general interest is to derive from (\ref{IntMainFull}) the effective dynamics of a charge subject to slowly varying external potentials and 
 coupled to the radiation field. Very crudely, one considers the subspace of $L^2(\BbbR^3; \BbbC^2)\otimes \Fock$ spanned by the ground states
  of $H(P)$ with $P\in \BbbR^3$ 
  and constructs the effective dynamics as an approximate solution to the full dynamics
  lying  close to that subspace. In principle this problem can be handled by space-adiabatic perturbation theory \cite{PST, Teuful}, which as one basic 
  input uses that $H(P)$ has a uniform spectral gap, i.e.,  for all $P\in \BbbR^3$, 
 \begin{align}
\inf \{\mathrm{spec}(H(P))\backslash \{E(P)\}\}-E(P)=:C_{\mathrm{g}}(P)\ge C_0 >0.\label{NRGap}
\end{align}  
This can be achieved by having a maximal velocity which is  strictly less than $1$, which means to choose $\gamma<1$. This choice amounts to  a small modification for low energies which is anyhow the domain of validity
 of our model. 
 
 If $\mph>0$ and $\gamma <1$ 
it is easily seen that $\Delta(P)\ge C_0 >0$
 uniformly. However this does not yet establish a spectral gap in the sense of  (\ref{NRGap}), because beyond the ground state there could be 
 other eigenvalues in the interval $[E(P), \Sigma(P)]$. In the literature there are two methods to count  the number of eigenvalues.
 One is through positive commutator,  Mourre type estimates and the other uses a pull through in order to estimate the overlap between the Fock vacuum and the 
 ground state. For sufficiently small $P$ both  methods yield the desired result. However, a uniform bound on the spectral gap  seems to be difficult
 to achieve by such techniques. Therefore we introduce a novel method based on  operator  monotonicity, which we learned from the masterly works of Lieb and Loss \cite{LL3, LL4},  together with  the min-max principle.
 
 Progressed so far, one still has to determine the degeneracy of the ground  state. For the non-relativistic Pauli-Fierz model this is discussed in \cite{HS}.
 Later on we learned  a very simple and general argument from M. Loss.  We reproduce  his result and show that it is applicable to the semi-relativistic
 Pauli-Fierz hamiltonian.
 
 We summarize our main result in  
 \begin{Thm}\label{UniformGap} 
Fix $0<\gamma<1$ and $0<\mph$ .  Then there exists $e_*>0$ independent of 
$P$,  such that, for all $e<e_*$ and $P\in\BbbR^3$, the following properties  hold. 
\begin{itemize} 
\item[{\rm (i)}] One has  
\begin{align*} 
 \Sigma(P)-E(P)\ge (1-\gamma)\mph-ec_1-\Od(e^2)>0 
\end{align*} 
for all $P\in \BbbR^3$, where $c_1$ and $\Od(e^2)$ are independent of 
$P$. 
In particular $E(P)$ is an eigenvalue. 
\item[{\rm (ii)}] One has  
\begin{align} 
\inf \big\{ \mathrm{spec}(H(P))\backslash \{E(P)\}\big\}-E(P)\ge (1-e 
c_2 -\gamma)\mph-ec_3-\Od(e^2) 
\end{align} 
for all $P$, where $c_2, c_3$ and $\Od(e^2)$ are independent of $P$.
\item[{\rm (iii)}]  $E(P)$ is exactly doubly degenerate. 
\end{itemize} 
\end{Thm}  

\begin{rem}
{\rm
We remark that the lowest energy $E(P)$ and a possible spectral gap are also of  importance,
 e.g., in scattering theory. We refer to \cite{FGS1, FGS2, Hiroshima} for the investigation of related models and to \cite{Chen}
 for $E(P)$ when the infrared cutoff is removed. $\diamondsuit$
}
\end{rem}

 \begin{flushleft}
 {\large {\bf Acknowledgements}} 
 \end{flushleft}
 We  would like to thank M. Loss for explaining to us how to use Kramers degeneracy in our context.
 This research is supported  by the DFG under the grant SP181/24.

\section{Self-adjointness}\label{SAGS} 
\subsection{Dirac operators} \label{DiracOps}
As a preliminary, we introduce two Dirac operators which will simplify  our 
study. 
 
Let us define a Dirac operator $D$ by 
\[ 
 D=\alpha\cdot (-\im \nabla_x+eA(x))+M\beta 
\] 
living in $L^2(\BbbR^3;\BbbC^4)\otimes \Fock$. This is essentially 
self-adjoint on $C_0^{\infty}(\BbbR^3; \BbbC^4)\otimes 
\Fock_{\mathrm{fin}}$ by the Nelson's commutator theorem \cite{ReSi2} with the test 
operator $-\Delta_x+\Hf$. Here 
\begin{align*} 
\Fock_{\mathrm{fin}}=\mathrm{Lin}\big\{a(f_1)^*\cdots a(f_n)^*\Omega,\, 
 \Omega\,|\, f_1(\cdot, \lambda_1), \dots, f_n(\cdot, \lambda_n)\in 
 C_0^{\infty}(\BbbR^3)\\\, \mbox{for all $\lambda_1,\dots, 
 \lambda_n\in \{1,2\}$ and $n\in \BbbN$}\big\}, 
\end{align*} 
where $\mathrm{Lin}\{\cdots\}$ means the linear span of the set 
 $\{\cdots\}$ and $\Omega$ is the Fock vacuum defined by $\Omega=1\oplus 
 0\oplus 0\oplus \cdots$. 
 We denote the  closure of $D$ by the same 
symbol. We note that  
\begin{align*} 
D^2&=T+M^2,\\ 
|D|&=\sqrt{T+M^2},
\end{align*}
where the self-adjoint operator $T$ is expressed as
\begin{align*} 
T&=\left(\begin{matrix} 
\big(\sigma\cdot (-\im\nabla_x+eA(x))\big)^2&0\cr 
0&\big(\sigma\cdot (-\im\nabla_x+eA(x))\big)^2&\cr 
\end{matrix} 
\right)\  
\end{align*} 
on $C_0^{\infty}(\BbbR^3; \BbbC^4)\otimes \Ffin$.
 
Next let us define the following Dirac operator 
\begin{align*} 
D(P)&=\alpha\cdot (P-\Pf+e\A)+M\beta\\ 
\end{align*}  
acting  in $\BbbC^4\otimes \Fock$. 
Again this is essentially self-adjoint on $\BbbC^4\otimes 
\Fock_{\mathrm{fin}}$ by the Nelson's commutator theorem with a test 
operator $\Pf^2+\Hf$. We denote its 
closure by the same symbol. 
Then one can easily observe that  
\begin{align*} 
\mathcal{U}D\mathcal{U}^*&=\int^{\oplus}_{\BbbR^3}D(P)\, \mathrm{d}P,\\ 
D(P)^2&=T(P)+M^2,\\ 
|D(P)|&=\sqrt{T(P)+M^2},
\end{align*}
where the action of the self-adjoint operator $T(P)$ is concretely given as 
\begin{align*}
T(P)&=\left(\begin{matrix} 
\big(\sigma\cdot (P-\Pf+e\A)\big)^2&0\cr 
0&\big(\sigma\cdot (P-\Pf+e\A)\big)^2&\cr 
\end{matrix} 
\right)\ 
\end{align*}  
 on $\BbbC^4\otimes \Ffin$.
 
\subsection{Definition of the Hamiltonians } 
Our definition of $H$ and $H(P)$ are as follow: 
\begin{align*} 
H&=\gamma |D|+\Hf,\\ 
H(P)&=\gamma |D(P)|+\Hf. 
\end{align*} 
In this paper we ocassionally identify a direct sum operator $A\oplus 
A$ with $A$ if no confusion occurs. Hence the above definitions mean  
that $H\oplus H=\gamma |D|+\Hf\oplus \Hf$ and $H(P)\oplus H(P)=\gamma |D(P)|+\Hf\oplus 
\Hf$. 
\subsection{Proof of Proposition \ref{SAA}} 
 
For each $\vphi\in C_0^{\infty}(\BbbR^3; \BbbC^4)\otimes 
\Fock_{\mathrm{fin}}$, one has  
\[ 
 \||D|\vphi\|^2=\la \vphi, D^2\vphi\ra\le \C \|(H_0+\one)\vphi\|^2. 
\] 
Since $C_0^{\infty}(\BbbR^3; \BbbC^4)\otimes 
\Fock_{\mathrm{fin}}$ is a core of $H_0$, one concludes that 
$\D(H_0)\subseteq \D(|D|)$. Also note that, for $D_0:=\alpha\cdot (-\im 
\nabla_x)+M\beta$, 
one has $\D(H_0)\subseteq \D(|D_0|)$. Let $H_{\mathrm{I}}$ be the 
interaction term given by \[ 
                           H_{\mathrm{I}}=|D|-|D_0|. 
\] 
By the above arguments, $\D(H_0)\subseteq \D(H_{\mathrm{I}})$ holds. Using 
the formula 
\begin{align} 
 |a|=\frac{1}{\pi}\int_0^{\infty}\mathrm{d}t\frac{1}{\sqrt{t}}\frac{a^2}{t+a^2},\label{Fundamental} 
\end{align} 
one has  
\begin{align} 
&|D|-|D_0|\nonumber\\ 
&=\frac{1}{\pi}\int_0^{\infty}\mathrm{d}t\sqrt{t}(t+D^2)^{-1}\Big\{2eA(x)\cdot 
 (-\im \nabla_x)+e^2A(x)^2+e\sigma\cdot B(x)\Big\}(t+D_0^2)^{-1},\label{DD0} 
\end{align} 
where $B(x)=\nabla_x\wedge A(x)$.
Observe that  
\begin{align*} 
&\|A_j(x)(-\im \partial_j)(t+D_0^2)^{-1}(H_0+\one)^{-1}\|\\ 
&\le\|A_j(x)(-\im \partial_j)|-\im\partial_j 
 |^{-1/2}(H_0+\one)^{-1}\|\||-\im \partial_j|^{1/2}(t+D_0^2)^{-1}\|\\ 
&\le \C t^{-3/4} 
\end{align*} 
for $j=1,2,3$, and  
\begin{align*} 
\|A(x)^2(H_0+\one)^{-1}\|&\le \C,\\ 
\|\sigma\cdot B(x)(H_0+\one)^{-1}\|&\le \C. 
\end{align*} 
Combining these with (\ref{DD0}), one obtains 
\begin{align} 
\|H_{\mathrm{I}}(H_0+\one)^{-1}\|&\le \C(e+e^2) 
 \int_0^{\infty}\mathrm{d}t\sqrt{t}(t+M^2)^{-1}\{t^{-3/4}+(t+M^2)^{-1}\}\nonumber\\ 
&\le \mathcal{O}(e).\label{InteractionEST} 
\end{align} 
Hence there exists $e_*$ such that
$\|H_{\mathrm{I}}(H_0+\one)^{-1}\|<1$  for all $e<e_*$. Now we can apply the 
Kato-Rellich theorem \cite{ReSi2} to obtain the assertion in Proposition \ref{SAA}.

\subsection{Proof of Proposition \ref{SAB}}
By (\ref{InteractionEST}), one has
\begin{align}
\|H_{\mathrm{I}}\psi\|\le \Od(e) \|(H_0+\one)\psi\|.\label{InteractionEST2}
\end{align}
For each $k_0\in \BbbR^3$, choose $\psi$ as  $\mathcal{U}\psi=|B_{
  \vepsilon, k_0}|^{-1/2}\chi_{B_{ \vepsilon, k_0}}\otimes \vphi$ where
  $\vphi\in \BbbC^2\otimes \Ffin$,  $\chi_S$ is the characteristic  function of the
  set $S$, $B_{\vepsilon, k_0}=\{k\in \BbbR^3\, |\,
  |k-k_0|<\vepsilon\}$ and $|B_{\vepsilon, k_0}|=4\pi \vepsilon^3/3$.
It follows from (\ref{InteractionEST2}) that 
\begin{align}
|B_{\vepsilon, k_0}|^{-1}\int_{B_{\vepsilon, k_0}}\mathrm{d}k\,
 \|H_{\mathrm{I}}(k)\vphi\|^2\le \Od(e^2)|B_{
 \vepsilon, k_0}|^{-1}\int_{B_{\vepsilon, k_0}}\mathrm{d}k\, \|(H_0(k)+\one)\vphi\|^2,
\end{align}
where $H_{\mathrm{I}}(P)=|D(P)|-|D_0(P)|$.
Since $H_{\mathrm{I}}(P)\vphi$ and $(H_0(P)+\one)\vphi$ are strongly
continuous in $P$, we can take the limit as $\vepsilon \downarrow 0$
and obtain that
\begin{align*}
\|H_{\mathrm{I}}(k_0)\vphi\| \le \Od(e)\|(H_0(k_0)+\one)\vphi\|.
\end{align*}
Since $k_0$ is arbitrary and $\BbbC^2\otimes \Ffin$ is a core of $H_0(P)$, we have
that $\|H_{\mathrm{I}}(P) (H_0(P)+\one)^{-1}\|\le \Od(e)$ for all $P$.
Now we can apply the Kato-Rellich theorem \cite{ReSi2} and obtain  the assertion in
the proposition. $\Box$

\section{Spectral properties}\label{ProofSpectrum} 
\subsection{Preliminaries} 
In this section, we will prove Theorem \ref{Spectrum}.  
To this end, we need some preliminaries. 
 
Let $j_1$ and $j_2$ be two localization functions on $\BbbR^3$ so that  
$j^2_1+j_2^2=1$ and $j_1$ is supported in a ball of radius  $R$. For 
each vector $f=f(k, \lambda)$ in $L^2(\BbbR^3\times \{1,2\})$, we define 
an operator 
$\J_i \ (i=1,2)$ by 
\[ 
 (\J_if)(k, \lambda)=j_i(-\im \nabla_k)f(k, \lambda). 
\] 
Now we define a linear operator $\J: L^2(\BbbR^3\times \{1,2\})\to 
L^2(\BbbR^3\times \{1,2\})\oplus L^2(\BbbR^3\times \{1,2\})$ 
 by  
\[ 
 \J f=\J_1 f\oplus \J_2 f 
\] 
for each $f\in L^2(\BbbR^3\times \{1,2\})$.  
 
Let $U$ be the natural isometry from $\Fock(L^2(\BbbR^3\times 
\{1,2\})\oplus L^2(\BbbR^3\times \{1,2\}))$ to $\Fock\otimes \Fock$ 
where $\Fock(L^2(\BbbR^3\times 
\{1,2\})\oplus L^2(\BbbR^3\times \{1,2\}))$ is the Fock space over $L^2(\BbbR^3\times 
\{1,2\})\oplus L^2(\BbbR^3\times \{1,2\})$, see Appendix \ref{SecondQ}. Concrete action of $U$ is 
given by 
\begin{align*} 
& Ua(f_1\oplus g_1)^*\dots a(f_n\oplus g_n)^*\Omega^{\oplus}\\ 
&=[a(f_1)^*\otimes 
 \one+\one \otimes a(g_1)^*]\dots [a(f_n)^*\otimes 
 \one+\one \otimes a(g_n)^*]\Omega\otimes \Omega  
\end{align*} 
where $\Omega^{\oplus}$ is the Fock vacuum in $\Fock(L^2(\BbbR^3\times 
\{1,2\})\oplus L^2(\BbbR^3\times \{1,2\}))$. 
The following  operator  
\[ 
 \BGamma(\J):=U\Gamma(\J) 
\] 
plays an  important role in our proof.  The
importance of $\BGamma(\J)$ was discovered by  Derezi\'nski and G\'erard \cite{DG1}.
 
In Appendix \ref{ProofLocalization}  we show the following formula. 
\begin{lemm}\label{Localization}{\rm (Localization formula)} 
Let  
\[ 
 H^{\otimes}(P)=\gamma \sqrt{\big\{\sigma\cdot(P-\Pf\otimes\one-\one\otimes \Pf+e\A\otimes 
 \one)\big\}^2+M^2}+\Hf\otimes\one+\one\otimes \Hf 
\] 
acting in $\BbbC^2\otimes \Fock\otimes \Fock$. Choose $e$ as $e<e_*$, 
 where $e_*$ is given in Proposition \ref{SAB}. 
Then, for all $\vphi\in 
 \BbbC^2\otimes \Fock_{\mathrm{fin}}\otimes \Fock_{\mathrm{fin}}$, one 
 obtains 
\begin{align*} 
\big|\big\la \vphi, (H(P)-\BGamma(\J)^*H^{\otimes}(P)\BGamma(\J))\vphi\big\ra\big|\le 
 \mathcal{O}(R^0)\|(H(P)+\one)\vphi\|^2, 
\end{align*} 
where $\mathcal{O}(R^0)$ is a function of $R$  vanishing as $R\to \infty$. 
\end{lemm} 
 Finally  we note the following lemma. 
 
\begin{lemm}\label{TensorSpec} 
One has  
\[ 
H^{\otimes}(P)\ge E(P)+\Delta(P)(\one-P_{\Omega}), 
\] 
where $P_{\Omega}$ is the orthogonal projection onto $\BbbC^2\otimes 
 \Fock \otimes \Omega$. 
\end{lemm} 
{\it Proof.} Remark the following natural identification, 
\[ 
 \BbbC^2\otimes \Fock\otimes\Fock =\Sumoplus\BbbC^2\otimes\Fock\otimes  
L^2(\BbbR^3\times \{1,2\})^{\otimes_{\mathrm{s}}n}.  
\] 
Set $\mathscr{H}_n=\BbbC^2\otimes\Fock\otimes  
L^2(\BbbR^3\times \{1,2\})^{\otimes_{\mathrm{s}}n}$. Each vector in 
$\vphi\in \mathscr{H}_n$ can be expressed as  
a $\BbbC^2\otimes \Fock$-valued symmetric function on $(\BbbR^3\times 
\{1,2\})^{\times n}$: 
\[ 
 \vphi=\vphi(k_1, \lambda_1, \dots, k_n, \lambda_n). 
\] 
Under this identification, the action of $H^{\otimes }(P)$ is given by 
\begin{align*} 
&(H^{\otimes }(P)\vphi)(k_1, \lambda_1, \dots, k_n, \lambda_n)\\ 
&=\Big(H\Big(P-\sum_{i=1}^nk_i\Big)+\sum_{i=1}^n\omega(k_i)\Big)\vphi(k_1, 
 \lambda_1, \dots, k_n, \lambda_n) 
\end{align*} 
for a suitable $\vphi\in \mathscr{H}_n$. Thus, using the triangle 
inequality $\omega(k_1+k_2)\le \omega(k_1)+\omega(k_2)$, 
 one has  
\begin{align*} 
\la \vphi, H^{\otimes }(P)\vphi\ra&=\sum_{\lambda_1, \dots, 
 \lambda_n=1,2}\int\mathrm{d}k_1\dots \mathrm{d}k_n\, \Big\la \vphi(k_1, 
 \lambda_1, \dots, k_n, \lambda_n),\\ 
 &\Big(H\Big(P-\sum_{i=1}^nk_i\Big)+\sum_{i=1}^n\omega(k_i)\Big)\vphi(k_1, 
 \lambda_1, \dots, k_n, \lambda_n)\Big\ra\\ 
&\ge (\Delta(P)+E(P))\|\vphi\|^2. 
\end{align*} 
For $n=0$, we have $H^{\otimes}(P)\restriction 
\mathscr{H}_0=H(P)$. Combining  the results, one reaches the assertion 
in the lemma. $\Box$ 
 
\subsection{Proof of Theorem \ref{Spectrum}} 
\subsubsection{Lower bound of $\Delta(P)$} 
In this subsubsection, we will show the following lower bound. 
\begin{Prop}\label{SpecLower} Choose $e<e_*$. Then one has that  
$\Sigma(P)-E(P)\ge \Delta(P).$ 
\end{Prop} 
{\it Proof.} 
For any $\lambda\in \mathrm{ess. spec}(H(P))$, we can find a sequence 
$\{\vphi_n\}_n$ 
such that $\|\vphi_n\|=1, \wlim \vphi=0$ and $\displaystyle \lim_{n\to 
\infty}\|(H(P)-\lambda)\vphi_n\|=0$. For each $n\in\BbbN$, one has  
\begin{align*} 
\la \vphi_n, H(P)\vphi_n\ra\ge \la \vphi_n, \BGamma(\J)^* 
 H^{\otimes}(P)\BGamma(\J)\vphi_n\ra 
-\mathcal{O}(R^0)\|(H(P)+\one)\vphi\|^2 
\end{align*} 
by Lemma \ref{Localization}. Thus using 
Lemma \ref{TensorSpec} one gets 
\begin{align} 
&\la \vphi_n, H(P)\vphi_n\ra\nonumber\\ 
&\ge 
E(P)+\Delta(P)-\Delta(P)\|P_{\Omega}\BGamma(\J)\vphi_n\|^2 
-\mathcal{O}(R^0)\|(H(P)+\one)\vphi_n\|^2.\label{AppEst} 
\end{align} 
 
First we will show that $\displaystyle \lim_{n\to \infty}\|P_{\Omega}\BGamma(\J)\vphi_n\|=0$. 
Remark that $\|P_{\Omega}\BGamma(\J)\vphi_n\|=\|\Gamma(\J_1)\vphi_n\|$. 
With $\Nf$  the number operator given by 
\[ 
 \Nf=\sum_{\lambda=1,2}\int_{\BbbR^3}\mathrm{d}k\, a(k, \lambda)^*a(k, 
 \lambda), 
\] 
we also remark that $\la  \vphi_n, \Nf\vphi_n\ra$ is uniformly bounded in 
$n$ because  
\[ 
 \la \vphi_n, \Nf \vphi_n\ra \le m_{\mathrm{ph}}^{-1}\la \vphi_n, 
 \Hf\vphi_n\ra\le \la \vphi_n, H(P)\vphi_n\ra. 
\] 
Thus $\|(\one-\chi_N(\Nf))\Gamma(\J_1)\vphi\|\le 
\|(\one-\chi_N(\Nf))\vphi_n\|=\mathcal{O}(N^0)$ holds where 
$\mathcal{O}(N^0)$ is a function of $N$, independent of $n$, vanishing as 
$N\to \infty$. Here $\chi_N(s)=1$ if $0\le s\le N$ and  $\chi(s)=0$ 
otherwise, moreover $\chi_N(\Nf)$ is defined by the functional calculus. 
On the other 
hand, $\chi_N(\Nf)(\Hf+\one)^{-1/2}\Gamma(\J_1)$ is compact for all 
$N$. Thus one finds that  
\begin{align*} 
&\|P_{\Omega}\BGamma(\J)\vphi_n\|^2\\ 
&\le 2 
 \|\chi_N(\Nf)\Gamma(\J_1)\vphi_n\|^2+2\|(\one-\chi_N(\Nf))\vphi_n\|^2\\ 
&=2\la \chi_N(\Nf)(\Hf+\one)^{-1/2}\Gamma(\J_1)^2\vphi_n, 
 (\Hf+\one)^{1/2}\vphi_n\ra+\mathcal{O}(N^0)\\ 
&=2\|\chi_N(\Nf)(\Hf+\one)^{-1/2}\Gamma(\J_1)^2\vphi_n\| 
 \|(\Hf+\one)^{1/2}(H(P)+\one)^{-1/2}\|\\ 
&\ \ \ \ \times \|(H(P)+\one)^{1/2}\vphi_n\| +\mathcal{O}(N^0). 
\end{align*} 
First we  take the limit $n\to \infty$. Then, by the compactness of the 
linear operator $\chi_N(\Nf)(\Hf+\one)^{-1/2}\Gamma(\J_1)$, the vector 
$\chi_N(\Nf)(\Hf+\one)^{-1/2}\Gamma(\J_1)^2\vphi_n$ converges to  $0$ 
strongly which implies that $\displaystyle \limsup_{n\to 
\infty}\|P_{\Omega}\BGamma(\J)\vphi_n\|\le \mathcal{O}(N^0)$. Then 
taking $N\to \infty$, one concludes that $\displaystyle \lim_{n\to \infty}\|P_{\Omega}\BGamma(\J)\vphi_n\|=0. $   
 
Taking the limit $n\to \infty$ in both side of (\ref{AppEst}), one finds 
\[ 
 \lambda\ge E(P)+\Delta(P)-\mathcal{O}(R^0)(\lambda+1)^2. 
\] 
Finally taking $R\to \infty$, one obtains the desired assertion. $\Box$

\subsubsection{Upper bound of $\Delta(P)$} 
We will complete our  proof of Theorem \ref{Spectrum} by showing the 
following upper bound. 
\begin{Prop}\label{SpecUpper} 
Choose $e$ as $e<e_*$. Then we have that $\Sigma(P)-E(P)\le \Delta(P)$. 
\end{Prop} 
{\it Proof.} For notational simplicity we set $\gamma=1$ in this proof. 
For each  $k_0\in \BbbR^3$, let us define  
\begin{align*} 
f_{\vepsilon, k_0}&=|B_{\vepsilon, k_0}|^{-1/2}\chi_{B_{\vepsilon, 
 k_0}},\\ 
B_{\vepsilon, k_0}&=\{k\in \BbbR^3\, |\, |k-k_0|\le \vepsilon\}, 
\end{align*} 
where $\chi_A$ is the characteristic function of the measurable set $A$ 
and $|A|$ means the Lebesgue measure of $A$. Choose a normalized vector 
$\vphi_{\vepsilon}\in \ran E_{\Delta}(H(P-k_0))$ with 
$\Delta=[-\vepsilon+z, z+\vepsilon], \ 
z=E(P-k_0)+\omega(k_0), \vepsilon>0$. Here for a self-adjoint operator $A$, 
$E_{\Delta}(A)$ stands for the spectral measure of $A$ for the 
interval $\Delta$. Let 
$a_{\lambda}(f)=\int_{\BbbR^3}\mathrm{d}k\, f(k)^* a(k, \lambda) 
$. 
We will show that $a_{\lambda}(f_{\vepsilon, 
  k_0})^*\vphi_{\vepsilon}/\|a_{\lambda}(f_{\vepsilon, 
  k_0})^*\vphi_{\vepsilon}\|$ 
is a Weyl sequence for $z$ as $\vepsilon\downarrow 0$. 
Applying the 
pull-through formula, one has  
\begin{align} 
&\big\la (H(P)-z)a_{\lambda}(f_{\vepsilon, k_0})^*\vphi_{\vepsilon}, 
 \psi\big\ra\nonumber\\ 
&=\int_{\BbbR^3}\mathrm{d}k\, f_{\vepsilon, k_0}(k)\Big\{ \big\la 
 (H(P-k)+\omega(k)-z)\vphi_{\vepsilon}, a(k, 
 \lambda)\psi\big\ra 
-\la \s(P)^*\vphi_{\vepsilon}, \psi\ra\Big\}\label{AdjPullThrough} 
\end{align} 
for each normalized  $\psi\in \BbbC^2\otimes \Fock_{\mathrm{fin}}$,  
where 
\[ 
 \s(P)=|D(P-k)|a(k, \lambda)-a(k, \lambda)|D(P)|. 
\] 
 As to the second term in the right 
hand side of (\ref{AdjPullThrough}), observe that  
\begin{align*} 
&\Big|\int_{\BbbR^3}\mathrm{d}k\, f_{\vepsilon, k_0}(k)\la 
 \s(P)^*\vphi_{\vepsilon}, \psi\ra\Big|\\ 
&\le \int_{\BbbR^3}\mathrm{d}k\, 
 f_{\vepsilon, k_0}(k)\|\s(P)^*\vphi_{\vepsilon}\|\|\psi\|\\ 
&\le \int_{\BbbR^3}\mathrm{d}k\, f_{\vepsilon, k_0}(k)\|\s(P)^*(H(P-k)+\one)^{-1}\|\|(H(P-k)+\one)\vphi_{\vepsilon}\|\\ 
&\le C \int_{\BbbR^3}\mathrm{d}k\, f_{\vepsilon, k_0}(k)(1+|k|)|F_0(k, \lambda)|\big(E(P-k_0)+1+\omega(k_0)+\Od(|k-k_0|)+\vepsilon \big) 
\end{align*}  
by Lemma \ref{AppB5} and (\ref{UpperEst12}) below, where  
\begin{align} 
F_x(k, \lambda)=e\frac{\chi_{\Lambda}(k)\vepsilon(k, 
  \lambda)}{\sqrt{2(2\pi)^3\omega(k)}}\ex^{-\im k\cdot x}. \label{FormFactor}
\end{align} 
 Clearly the right hand side of the above inequality 
converges to $0$ as $\vepsilon \downarrow 0$ because $f_{\vepsilon, 
k_0}$ weakly converges to $0$ in $L^2(\BbbR^3)$. Next we will estimate the first term in 
the right hand side of (\ref{AdjPullThrough}). One has  
\begin{align} 
&\Big|\int_{\BbbR^3}\mathrm{d}k\, f_{\vepsilon, k_0}(k)\big\la 
 (H(P-k)+\omega(k)-z)\vphi_{\vepsilon}, a(k, 
 \lambda)\psi\big\ra\Big|\nonumber\\ 
&\le \Big[\int_{\BbbR^3}\mathrm{d}k\, f_{\vepsilon, 
 k_0}(k)^2\big\|(\Nf+2)^{1/2}(H(P-k)+\omega(k)-z) 
\vphi_{\vepsilon}\big\|^2\Big]^{1/2}\label{UpperEst1}\\ 
&\  \ \ \times \Big[\int_{\BbbR^3}\mathrm{d}k\, \big\|(\Nf+2)^{-1/2}a(k, 
 \lambda)\psi\big\|^2\Big]^{1/2}\label{UpperEst2} 
\end{align}  
The term (\ref{UpperEst2}) is less than $\|\psi\|^2(=1)$ because  
\begin{align*} 
\int_{\BbbR^3}\mathrm{d}k\, \big\|(\Nf+2)^{-1/2}a(k, 
 \lambda)\psi\big\|^2&\le\int_{\BbbR^3}\mathrm{d}k\, \la a(k, 
 \lambda)\psi, a(k, \lambda)(\Nf+\one)^{-1}\psi\ra\\ 
&=\la \psi, \Nf(\Nf+\one)^{-1}\psi\ra \\ 
&\le\|\psi\|^2. 
\end{align*} 
As to  the term (\ref{UpperEst1}) we need a lengthy calculation below. 
 Note that, since $\Nf+2\le 
\mph^{-1}H(P-k)+2$, one has  
\begin{align} 
&\big\|(\Nf+2)^{1/2}(H(P-k)+\omega(k)-z)\vphi_{\vepsilon}\big\|\nonumber\\ 
&\le  
 C\big\|(H(P-k)+2)^{1/2}(H(P-k)+\omega(k)-z)\vphi_{\vepsilon}\big\|\nonumber\\ 
&\le  
 C\big\|(H(P-k)+\omega(k)-z)^{3/2}\vphi_{\vepsilon}\big\|\label{UpperEst3}\\ 
&\ \ \ + C|2-z+\omega(k)|^{1/2} 
 \big\|(H(P-k)+\omega(k)-z)\vphi_{\vepsilon}\big\|.\label{UpperEst4} 
\end{align} 
Note that   
\begin{align*} 
&H(P-k)+\omega(k)-z\\ 
&=\big(H(P-k_0)+\omega(k_0)-z\big)+\big(H(P-k)-H(P-k_0)+\omega(k)-\omega(k_0)\big). 
\end{align*} 
Thus one has  
\begin{align} 
&\big\|(H(P-k)+\omega(k)-z)\vphi_{\vepsilon}\big\|\nonumber\\ 
&\le  
 \big\|(H(P-k_0)+\omega(k_0)-z))\vphi_{\vepsilon}\big\|+\big\|(H(P-k)-H(P-k_0))\vphi_{\vepsilon}\big\|\nonumber\\ 
&\ \ \ +|\omega(k)-\omega(k_0)|\nonumber\\ 
&\le \vepsilon+\|(|D(P-k)|-|D(P-k_0)|)\vphi_{\vepsilon}\|+|\omega(k)-\omega(k_0)|.\label{UpperEst11} 
\end{align} 
We will show that 
\begin{align} 
\big\|(|D(P-k)|-|D(P-k_0)|)(H(P-k_0)+\one)^{-1}\big\|\le \mathcal{O}(|k-k_0|).\label{UpperEst5} 
\end{align}  
To see this, we  just note that, by (\ref{Fundamental}), 
\begin{align*} 
& |D(P-k)|-|D(P-k_0)|\\ 
&=\frac{1}{\pi}\int_{M^2}^{\infty}\mathrm{d}t\, 
 \sqrt{t}(t+\hat{D}(P-k)^2)^{-1}\Big\{2(-k+k_0)\cdot\Pf+k^2-k_0^2 
+2e (-k+k_0)\cdot \A 
\Big\}\\ 
&\ \ \ \times (t+\hat{D}(P-k_0)^2)^{-1}. 
\end{align*} 
Hence, by Lemma \ref{AppB2}, one obtains 
\begin{align} 
&\big\|(|D(P-k)|-|D(P-k_0)|)(H(P-k)+\one)^{-1}\big\|\nonumber\\ 
&\le  \frac{1}{\pi}\int_{M^2}^{\infty}\mathrm{d}t\, \sqrt{t}t^{-1}\Big\|\Big\{2(-k+k_0)\cdot\Pf+k^2-k_0^2 
+2e (-k+k_0)\cdot \A 
\Big\}(\Hf+\one)^{-1}\Big\|\nonumber \\ 
&\ \ \ 
 \times\underbrace{\big\|(\Hf+\one)(t+\hat{D}(P-k_0)^2)^{-1}(H(P-k_0)+\one)^{-1}\big\|}_{\le C (t^{-1}+t^{-3/2}+t^{-2})\ \mbox{by Lemma \ref{AppB2}}}\nonumber\\ 
&\le \mathcal{O}(|k-k_0|)  \label{UpperEst6} 
\end{align}  
which implies 
\begin{align} 
\big\|(H(P-k)+\omega(k)-z)\vphi_{\vepsilon}\big\|\le \mathcal{O}(|k-k_0|)+\vepsilon \label{UpperEst12} 
\end{align} 
by (\ref{UpperEst11}). 
As a consequence, the term (\ref{UpperEst4}) is estimated as  
\begin{align} 
(\ref{UpperEst4}) \le  (\mathcal{O}(|k-k_0|)+\vepsilon) |2-z+\omega(k)|^{1/2}. \label{UpperEst7} 
\end{align} 
To estimate (\ref{UpperEst3}), observe that  
\begin{align} 
&\big\|(H(P-k)+\omega(k)-z)^{3/2}\vphi_{\vepsilon}\big\|^2\nonumber\\ 
&\le \big\|(H(P-k)+\omega(k)-z)^2\vphi_{\vepsilon}\big\|\underbrace{\big\|(H(P-k)+\omega(k)-z)\vphi_{\vepsilon}\big\|}_{\le 
 \mathcal{O}(|k-k_0|)+\vepsilon \ \mbox{by (\ref{UpperEst12})}}.\label{UpperEst8}  
\end{align} 
Since $\|(H(P-k)+\omega(k)-z)^2(H(P-k_0)+1))^{-2}\|$ is uniformly 
bounded  for  $k\in B_{\vepsilon, k_0}$ by Lemma \ref{AppB4}, we obtain 
\begin{align} 
(\ref{UpperEst3})\le \mathcal{O}(|k-k_0|)+C\vepsilon.\label{UpperEst9} 
\end{align}  
Collecting (\ref{UpperEst7}) and (\ref{UpperEst9}), we arrive at 
\begin{align*} 
&\int_{\BbbR^3}\mathrm{d}k\, f_{\vepsilon, 
 k_0}(k)^2\big\|(\Nf+\one)^{1/2}(H(P-k)+\omega(k)-z) 
\vphi_{\vepsilon}\big\|^2\\ 
&\le  \int_{\BbbR^3}\mathrm{d}k\, f_{\vepsilon, k_0}(k)^2 
 (\mathcal{O}(|k-k_0|)+\vepsilon)\\ 
&=\mathcal{O}(\vepsilon^0). 
\end{align*}  
This means that  
\begin{align} 
\big\|(H(P)-z)a_{\lambda}(f_{\vepsilon, 
 k_0})^*\vphi_{\vepsilon}\big\|\le \mathcal{O}(\vepsilon^0).\label{UpperEst10} 
\end{align} 
Set $\Psi_{\vepsilon}=a_{\lambda}(f_{\vepsilon, 
k_0})^*\vphi_{\vepsilon}/\|a_{\lambda}(f_{\vepsilon, 
k_0})^*\vphi_{\vepsilon}\|$. (Note that, by the CCRs, 
$\|a_{\lambda}(f_{\vepsilon, 
k_0})^*\vphi_{\vepsilon}\|^2=\|f_{\vepsilon, 
k_0}\|^2\|\vphi_{\vepsilon}\|^2+\|a_{\lambda}(f_{\vepsilon, 
k_0})\vphi_{\vepsilon}\|^2\ge 1$.) Then one can easily see that 
$\Psi_{\vepsilon}$  weakly converges to $0$ as $\vepsilon\downarrow 0$ and, by 
(\ref{UpperEst10}), $\lim_{\vepsilon\downarrow 
0}\|(H(P)-z)\Psi_{\vepsilon}\|=0$. 
Hence $\{\Psi_{\vepsilon}\}$ is a Weyl sequence. Thus 
$z=E(P-k_0)+\omega(k_0)\in \mathrm{ess. spec}(H(P))$. 
Since $k_0$ is arbitrary, one has the desired assertion in the 
proposition. $\Box$

\section{ Degenerate eigenvalues}\label{SectionDege} 
\subsection{Abstract Kramers' degeneracy theorem} 
The following lemma is well-known as the {\it Kramers' degeneracy 
theorem} which plays a central role in this section. 
 
\begin{lemm}\label{Kramers}{\rm (Abstract Kramers' degeneracy  theorem)} 
Let $\vartheta$ be an antiunitary operator with $\vartheta ^2=-\one$. (In applications  $\vartheta $ is mostly  the time reversal 
 operator.) Let $H$ be a 
 self-adjoint operator.   Assume that $H$ commutes with $\vartheta$. Then each 
 eigenvalue of $H$ is at least doubly degenerate. 
\end{lemm} 
{\it Proof.} Let $\mu$ be an eigenvalue of $H$ and $x$ be a 
corresponding eigenvector. Then, by the polarization identity,  one has $\la x, \vartheta x\ra=\la \vartheta \vartheta x, 
\vartheta x\ra=-\la x, \vartheta x\ra$ which means $x$ and $\vartheta x$ are orthogonal to each 
other. On the other hand, $\vartheta x$ is an  eigenvector with eigenvalue 
$\mu$ too. Hence $\mu$ is degenerate. $\Box$ 
\subsection{Reality preserving operators and degenerate eigenvalues} \label{RealPres}
 
Recall that the Hamiltonian  $H(P)$ is living in $\BbbC^2\otimes \Fock$. 
Each vector $\vphi\in\BbbC^2\otimes \Fock$ has the following expression: 
\begin{align*} 
\vphi&=\vphi_1\oplus \vphi_2,\\ 
\vphi_i&=\Sumoplus \vphi_i^{(n)}(k_1,\lambda_1,\dots, k_n,\lambda_n),\ i=1,2,
\end{align*} 
under the identification $\BbbC^2\otimes \Fock=\Fock\oplus \Fock$.
For each $\vphi\in \BbbC^2\otimes \Fock$, set  
\begin{align*} 
J\vphi&=j\vphi_1\oplus j\vphi_2,\\ 
j\vphi_i&=\Sumoplus \overline{\vphi}_i^{(n)}(k_1,\lambda_1,\dots, k_n,\lambda_n),\ i=1,2. 
\end{align*} 
We say that a linear operator $A$ on $\Fock$ {\it preserves the reality 
w.r.t. $j$} if $A$ commutes with $j$. 
 
Since $a(k, \lambda)$ acts by 
\[ 
 a(k,\lambda)\vphi_i=\Sumoplus \sqrt{n+1}\vphi_i^{(n+1)}(k,\lambda, k_1, 
 \lambda_1, \dots, k_n, \lambda_n), 
\] 
one has  
$ 
 ja(k,\lambda)=a(k,\lambda)j$ and $ ja(k,\lambda)^*=a(k,\lambda)^*j 
$ 
which imply 
\begin{align} 
j\Pf&=\Pf j, \label{J1}\\ 
j\A&=\A j,\label{J2}\\ 
j\B&=-\B j,\label{J3}\\ 
jH_{\mathrm{f}}&=H_{\mathrm{f}}j,\label{J4} 
\end{align} 
that is, $\Pf, \A, \im\B$ and $\Hf$ preserve the reality w.r.t. $j$.  (Here 
$\B=\nabla\wedge \A$.)

\begin{Prop}\label{AtLeastTwo}  
Let the time reversal operator $\vartheta$ be given by 
\[ 
 \vartheta=\sigma_2 J. 
\]  
For all $P$ and $e$,  we obtain that  
\[ 
 \vartheta H(P)=H(P)\vartheta. 
\] 
Thus, by Lemma \ref{Kramers},  each eigenvalue of $H(P)$ is \underline{at least} doubly degenerate. 
\end{Prop} 
 {\it Proof. } Since $\Hf $ commutes with $\vartheta$ by (\ref{J4}), it sufficies to  show that $|D(P)|$ commutes with $\vartheta$.
 Our basic idea is simple.
 Noting the fact $\vartheta\sigma_i=-\sigma_i\vartheta$ for $i=1,2,3$, we can easily see that  
$\vartheta$ commutes with $|D(P)|^2(=(P-\Pf+e\A)^2+e\sigma\cdot \B+M^2)$ on $\BbbC^2\otimes \Ffin$ by (\ref{J1})-(\ref{J3}). As a consequence
we could expect that  
$|D(P)|(=\sqrt{(P-\Pf+e\A)^2+e\sigma\cdot \B+M^2})$  also commutes with $\vartheta$. 

Unfortunately since we do not know whether the subspace $\BbbC^2\otimes \Ffin$ is a core of $D(P)^2$ or not, 
the above arguments are somehow formal. However we can rigorize the arguments as follow.
To clarify the $M$ dependence, we write $D(P)$ as $D_M(P)$ in this proof.  Let $\tilde{\vartheta}=\vartheta\oplus \vartheta$
acting in $\BbbC^4\otimes \Fock$. Then we see that $\alpha_i \tilde{\vartheta}=-\tilde{\vartheta}\alpha_i$ and $\beta\tilde{\vartheta}=\tilde{\vartheta}\beta$
which imply  $\tilde{\vartheta}D_M(P)\tilde{\vartheta}^{-1}=-D_{-M}(P)$ on $\BbbC^4\otimes \Ffin$. Since we have already seen that $\BbbC^4\otimes 
\Ffin$ is a core of 
$D_M(P)$ in section \ref{DiracOps}, this equality holds as an operator equality. Hence, by the functional calculus,  one has $\tilde{\vartheta} f(D_M(P))\tilde{\vartheta}^{-1}=f(-D_{-M}(P))$, where $f$ is real-valued.  In the case where $f(s)=\sqrt{s^2}$, we have $ f(-D_{-M}(P))=f(D_{M}(P))$
 because $D_{-M}(P)^2=D_M(P)^2$ by the anticommutativity between  $M\beta$ and $D_{M=0}(P)$. Now one can conclude that
 $\tilde{\vartheta}|D_M(P)|\tilde{\vartheta}^{-1}=|D_M(P)|$  holds as an operator equality. $\Box$

\subsection{Comments on related models} 
The arguments in this section  are  applicable to other models,  e.g., 
\begin{align*} 
H_{\mathrm{NR}, V}&=   \frac{1}{2M}\big(\sigma\cdot (-\im \nabla_x+ e A(x))\big)^2+\Hf+V(x),\\ 
H_{\mathrm{NR}}(P)&=\frac{1}{2M}\big(P-\Pf +e  A(0)\big)^2+\frac{e}{2M}
\sigma\cdot B(0)+\Hf\\
H_V&=\sqrt{(-\im \nabla_x+eA(x))^2+e\sigma\cdot B(x)+M^2}+V(x)+\Hf 
\end{align*} 
with $V(x)=V(-x)$. As regards to $H_{\mathrm{NR}}(P)$,  most of the  arguments 
 of section  \ref{RealPres} are valid. However, for  
$H_{\mathrm{NR}, V}$ and $H_V$, we have to change the 
definition of $j$. $H_{V}$ is acting in the Hilbert space $L^2(\BbbR^3; 
\BbbC^2)\otimes \Fock$. Each vector $\vphi$ in $L^2(\BbbR^3; 
\BbbC^2)\otimes \Fock$ has the form 
\begin{align*} 
\vphi&=\vphi_1\oplus \vphi_2,\\ 
\vphi_i&=\Sumoplus\vphi_i^{(n)}(x;k_1, \lambda_1, \dots, 
 k_n, \lambda_n), \ \ i=1,2. 
\end{align*} 
In this case, we define $j$ as  
\[ 
 j\vphi_i^{(n)}=\Sumoplus \overline{\vphi}_i^{(n)}(-x; k_1, \lambda_1, 
 \dots, k_n, \lambda_n),  \ \ i=1,2. 
\] 
Then one can check that  
\begin{align*} 
j(-\im \nabla_x)&=(-\im \nabla_x)j,\\ 
jA(x)&=A(x)j,\\ 
j(\im B(x))&=(\im B(x))j,\\ 
jV(x)&=V(x)j,\\ 
j\Hf&=\Hf j, 
\end{align*} 
namely, all these operators preserve the reality w.r.t. this new  $j$. 
Hence defining the time reversal operator as  $\vartheta=\sigma_2 J$, 
one can see that $H_V$ commutes with  $\vartheta$. Thus using the 
abstract Kramers' degeneracy theorem, one concludes that each eigenvalue 
of $H_V$ is at least doubly degenerate. A  similar modification applies  to 
$H_{\mathrm{NR}, V}$.

\section{Energy inequalities} 
To make sure that there is no further eigenvalues close to  $E(P)$ we will find self-adjoint operators $L_+(P)$ and $L_-(P)$
 such that 
 \begin{align}
 L_-(P)\le H(P)\le L_+(P).
 \end{align}
 $L_-(P), L_+(P)$ are given below.  They can be easily diagonalized. The min-max principle allows us to obtain bounds as, e.g., 
 \begin{align*}
 \Sigma(P)-E(P)\ge \Sigma(L_-(P))-E(L_+(P))
 \end{align*}
 and more precise information, since the spectrum of $L_{\pm}(P)$ is available,  see section \ref{ProofUniGap} for details.
 (Here, for a self-adjoint operator $T$, $\Sigma(T)=\inf \mathrm{ess. spec}(T)$ and $E(T)=\inf \mathrm{spec}(T)$. )
 
\begin{Prop}{\rm (Lower bound)}\label{LowerEnergyOp} 
For any $0<\gamma <1, 0\le \mph$ and $P\in\BbbR^3$,  one has 
\begin{align*} 
H(|P|u)\ge  L_-(P) 
\end{align*}
with 
\begin{align}
L_-(P)=\gamma \sqrt{P^2+M^2}+(1-\gamma-eC_1)\Hf-eC_2\label{LEO} 
\end{align} 
for suitable constants $C_1, C_2>0$ which are independent of $e$ 
and $P$, where $u=(1,0,0)$. 
\end{Prop} 
{\it Proof.} {\bf STEP 1.} Let $H_{\mathrm{SL}}(P)$ be the 
spinless Hamiltonian. In this step, we will show the following 
operator inequality by developing the method in \cite{LL4}: 
\begin{align} 
&H_{\mathrm{SL}}(|P|u)\nonumber \\ 
&\ge \gamma 
\sqrt{P^2+M^2} 
+(1-\gamma-eC)\Hf-eC
\label{SpinlessLower} 
\end{align}  
with a strictly positive constant $C$ independent of $e$ and $P$.  
Clearly  
\begin{align*} 
(|P|u-\Pf+e \A)^2\ge (|P|-P_{\mathrm{f}1}+e\A_1)^2. 
\end{align*}  
Thus by the operator monotonicity of the square root (Lemma \ref{OpMono}), one has  
\begin{align*} 
H_{\mathrm{SL}}(|P|u)\ge \gamma \sqrt{(|P|-P_{\mathrm{f}1}+e\A_1)^2+M^2}+\Hf. 
\end{align*}  
 
 Let $f(s)=\sqrt{s^2+M^2},\ s\in \BbbR$.  By Taylor's theorem, one has 
 \begin{align*}
 f(|P|+s)=f(|P|)+f'(|P|)s+\int_0^1\mathrm{d}t\, (1-t) f''(|P|+ts) s^2
 \end{align*}  
 with $f'(s)=s/\sqrt{s^2+M^2}$ and $f''(s)=M^2/(s^2+M^2)^{3/2}$.
Applying the functional calculus, we have the following  operator equality
\begin{align}
\sqrt{(|P|-P_{\mathrm{f1}}+eA(0)_1)^2+M^2}
=\sqrt{P^2+M^2}+\frac{|P|}{\sqrt{P^2+M^2}}(-P_{\mathrm{f1}}+eA(0)_1)\nonumber\\
+\int_0^1\mathrm{d}t\, (1-t) f''\big(|P|+t(-P_{\mathrm{f}1}+eA(0)_1)\big)
 (-P_{\mathrm{f}1}+eA(0)_1)^2.\label{Taylor}
\end{align}
Since the last term in (\ref{Taylor} ) is a positive operator, one obtains
\begin{align*}
&\sqrt{(|P|-P_{\mathrm{f1}}+eA(0)_1)^2+M^2}\\
&\ge \sqrt{P^2+M^2}+\frac{|P|}{\sqrt{P^2+M^2}}(-P_{\mathrm{f1}}+eA(0)_1)\\
&\ge \sqrt{P^2+M^2}-\Hf-\|\omega^{-1/2}F_{01}\|(\Hf+\one)
\end{align*}
by the standard bounds $|P_{\mathrm{f}1}|\le \Hf$ and $eA(0)_1\ge -\|\omega^{-1/2}F_{01}\|(\Hf+\one)$.
This proves   (\ref{SpinlessLower}). 
 
{\bf STEP 2.} We will show that  
\begin{align} 
\pm (H_{\mathrm{SL}}(|P|u)-H(|P|u))\le \frac{3\pi}{M} 
\|(1+\omega^{-1/2})|k||F_0|\|(\Hf+\one).\label{DiffSpin} 
\end{align} 
To this end, we simply note that, by (\ref{Fundamental}),  
\begin{align*} 
&H_{\mathrm{SL}}(|P|u)-H(|P|u)\\ 
&=-\frac{1}{\pi}\int_{M^2}^{\infty} \mathrm{d}s\, 
\sqrt{s-M^2}\big(s+(|P|u-\Pf+e \A)^2\big)^{-1}e\sigma\cdot \B 
(s+\hat{D}(|P|u)^2)^{-1},
\end{align*}  
where $\hat{D}(P)=D(P)-M\beta$.
Noting the facts $\|e\sigma\cdot \B(\Hf+\one)^{-1/2}\|\le 
6\|(1+\omega^{-1/2})|k||F_0|\|$ and (\ref{HfDHf}) in the proof of Lemma \ref{AppB2}, one can see that  
$\|(H_{\mathrm{SL}}(P)-H(P))(\Hf+\one)^{-1/2}\|\le 3\pi 
\|(1+\omega^{-1/2})|k 
||F_0|\|/M$. Now (\ref{DiffSpin}) is obtained. 
 
{\bf STEP 3.} (Proof of Proposition \ref{LowerEnergyOp}) 
From (\ref{SpinlessLower}) and (\ref{DiffSpin}) it follows that  
\begin{align*} 
H(|P|u)&=H_{\mathrm{SL}}(|P|u)+\big(H(|P|u)-H_{\mathrm{SL}}(|P|u)\big)\\
&\ge H_{\mathrm{SL}}-eC(\Hf+\one)  \\
&\ge \gamma \sqrt{P^2+M^2}+(1-\gamma-eC_1)\Hf-eC_2.
\end{align*}
This proves the desired assertion in the proposition. $\Box$  
 \medskip\\

Before we proceed, we remark the following. Let $SO(3)$ be the rotation group. Then there exists a unitary representation $\pi$
 of $SO(3)$  such that 
 \begin{align}
 \pi_g H(P)\pi_g^{-1}=H(g^{-1} P)  \label{Rotation}
 \end{align}
 for all $g\in SO(3)$ and $P\in \BbbR^3$, see e.g., \cite{Spohn2}.
 Thus $E(P)$ is a radial function in $P$.
 
Since $E(P)$ is rotationally symmetric in $P$, one has an immediate corollary.

\begin{coro} \label{EnLower}Choose $\gamma<1$ and $e$ sufficiently small as  $e<e_*$.
One has  
\begin{align*} 
E(P)\ge \gamma \sqrt{P^2+M^2}-eC_1
\end{align*} 
for all $P\in \BbbR^3$, where $C_1$ is independent of $e, P$. 
\end{coro}

\begin{Prop}\label{UpperE}{\rm (Upper bound)} One obtains 
\begin{align*} 
H(|P|u)\le L_+(P)
\end{align*}
with 
\begin{align}
L_+(P)
=\gamma 
&\Big[(|P|u-\Pf)^2+2|P|(\Hf+\|\omega^{-1/2}|F_0|\|)\nonumber\\ 
&+4(\Hf+\one)\Pf^2+\|(1+\omega^{-1/2})|F_0|\|^2 
+\|(1+\omega^{-1/2})|F_0|\|^2(\Hf+\one)\nonumber\\ 
&+\Hf+\||k|^{1/2}|F_0|\|^2+M^2 
\Big]^{1/2}+\Hf.\label{UpperEnergyOp} 
\end{align} 
for all $P$. 
\end{Prop}  
{\it Proof.} Observe that  
\begin{align*} 
D(|P|u)^2= (|P|u-\Pf)^2+2 (|P|u-\Pf)\cdot e \A+e^2 \A^2+e \sigma\cdot \B+M^2. 
\end{align*}  
Using the fundamental inequalities in Appendix A, one has  
\begin{align*} 
|P|e\A_1 &\le |P| (\Hf+\|\omega^{-1/2}|F_0|\|),\\ 
e\A\cdot P_{\mathrm{f}}&\le 
2(\Hf+\one)\Pf^2+\frac{1}{2}\|(1+\omega^{-1/2})|F_0|\|^2,\\ 
e^2 \A^2&\le \|(1+\omega^{-1/2})|F_0|\|^2(\Hf+\one),\\ 
e\sigma\cdot \B&\le \Hf+\||k|^{1/2}|F_0|\|^2. 
\end{align*} 
Applying the operator monotonicity of the square root (Lemma \ref{OpMono}), one concludes 
(\ref{UpperEnergyOp}). $\Box$ 
\medskip\\ 
 
By the above operator inequality, we immediately obtain 
\begin{align*} 
&\la \eta_{\uparrow}\otimes \Omega, H(|P|u)\eta_{\uparrow}\otimes \Omega\ra\\ 
&\le \gamma 
\sqrt{|P|^2+2|P|\|\omega^{-1/2}|F_0|\| 
+2\|(1+\omega^{-1/2})|F_0|\|^2+\||k|^{1/2}|F_0|\|^2+M^2},
\end{align*}  
where $\eta_{\uparrow}=(1,0)$.
Thus taking the rotational invariance of $E(P)$ in $P$  into 
consideration,  
one has a 
following corollary.

\begin{coro}\label{EnUpper} 
One has  
\begin{align} 
E(P)\le \gamma \sqrt{(|P|+e C_3)^2+M^2+e^2C_4}\label{EnergyUpper} 
\end{align}  
for all $P$, where $C_3$ and $C_4$ are independent of $P$ and $e$. 
\end{coro}

\section{Proof of Theorem \ref{UniformGap}}\label{ProofUniGap}
\subsection{Proof of Theorem \ref{UniformGap} (i)}  
 For $a\in \BbbR^d$, $\|a\|_{\BbbR^d}$ means the standard 
norm in $\BbbR^d$. Then one has, for example, $\omega(k)=\|(k, 
\mph)\|_{\BbbR^4}=\|(|k|, \mph)\|_{\BbbR^2}$. Applying the triangle 
inequality and Corollary \ref{EnLower}, one gets 
\begin{align*} 
&E(P-k)+\omega(k)\\ 
&\ge \gamma \|(|P-k|, M)\|_{\BbbR^2}+ 
 \|(|k|, 
\mph)\|_{\BbbR^2}-e C\\ 
&\ge \gamma \|(|P-k|, M)\|_{\BbbR^2}+\gamma \|(|k|, 
\mph)\|_{\BbbR^2}+(1-\gamma)\|(|k|, \mph)\|_{\BbbR^2}-eC\\ 
&=\gamma \|(P-k, M)\|_{\BbbR^4}+\gamma \|(k, 
\mph)\|_{\BbbR^4}+(1-\gamma)\|(k, \mph)\|_{\BbbR^4}-e C\\ 
&\ge  \gamma \|(P, M+\mph)\|_{\BbbR^4}+(1-\gamma)\mph-eC.   
\end{align*}  
On the other hand, since $\|\omega^{-1/2}|F_0|\|^2=\Od(e^2)$ etc., one has, 
by Corollary \ref{EnUpper}, that  
\begin{align*} 
E(P)&\le \gamma \|(|P|+eC_3, \sqrt{M^2+\Od(e^2)})\|_{\BbbR^2}\\ 
&\le \gamma \|(|P|+eC_3, M)\|_{\BbbR^2}+\gamma \|(0, 
\sqrt{M^2+\Od(e^2)}-M)\|_{\BbbR^2}\\ 
&\le \gamma \|(P, M)\|_{\BbbR^4}+eC_3+\Od(e^2).  
\end{align*}  
Thus the desired assertion in the lemma follows.  $\Box$

\subsection{Proof of Theorem \ref{UniformGap} (ii) and (iii)} 
We denote  the  infinimum  of $\mathrm{spec}(L_-(P))$ and 
$\mathrm{ess. spec}(L_-(P))$ by  $\EU(P)$ and $\Sigma_-(P)$ 
respectively. Clearly one has  
\begin{align*} 
\EU(P)&=\gamma \sqrt{P^2+M^2}-e C
,\\ 
\Sigma_-(P)&=\gamma \sqrt{P^2+M^2}+(1-\gamma -eC_1) \mph-eC_2.
\end{align*} 
Let $\EO(P)$ be the function of $P$ which is appearing on  the right 
hand side of (\ref{EnergyUpper}). Then using  similar arguments as  in 
the proof of Theorem \ref{UniformGap} (i), one sees that  
\begin{align} 
0\le \EU(P)-\EO(P)\le C'e+\Od(e^2)\label{UODiff}  
\end{align} 
with   $e<e_*$. (Here it should be noted that $e_*$ is independent of $P$.) 
Thus taking the fact $\EU(P)\le E(P) \le \EO(P)$ into consideration, 
one has  
\begin{align} 
0\le E(P)-\EU(P)\le C' e+\Od(e^2)\label{EUDiff} 
\end{align} 
 for $e<e_*$, which means $E(P)$ is close to $\EU(P)$ uniformly in $P$. 
Also we note that  
\begin{align} 
\Sigma_-(P)-\EU(P)\ge (1-eC-\gamma)\mph. \label{LGap} 
\end{align} 
 
Let $E_1(P)$ be the first excited eigenvalue of $H(P)$ (or possibly be 
$\Sigma(P)$ if there is no  such excited state.) Then by the operator 
inequality (\ref{LEO}) and the min-max principle \cite{ReSi4}, one has  
\begin{align*} 
E_1(|P|u)\ge \Sigma_-(P). 
\end{align*}  
(Note that, by Proposition \ref{AtLeastTwo}, $E(P)$ is always degenerate.) 
With the help of (\ref{Rotation}), one sees that $E_1(P)$ is radial and 
\begin{align} 
E_1(P)\ge \Sigma_-(P). \label{MinMaxResult} 
\end{align}  
for all $P\in \BbbR^3$.
Thus, combining this with (\ref{EUDiff}), we arrive at  
\begin{align*} 
E_1(P)-E(P)&\ge \Sigma_-(P)-\EO(P)\\ 
&\ge \Sigma_-(P)-\EU(P)+(\EU(P)-\EO(P))\\ 
&\ge (1 -e C_1-\gamma)\mph-C'e-\Od(e^2)   
\end{align*}  
 for $e<e_*$. This proves (ii) in the theorem.

For a self-adjoint operator $A$, let $E_K(A)$ be its spectral measure 
for the interval $(-\infty, K)$ and let $P_{\mathrm{pp}}(A)$ be the 
projection onto the linear space spanned by all eigenstates. Since, by Proposition 
\ref{LowerEnergyOp},  one 
has the operator inequality $L_-(P)\le H(|P|u)$, 
 the following property holds,
\begin{align*} 
\mathrm{tr} P_{\mathrm{pp}}(H(|P|u)) E_{\Sigma_-(P)}(H(|P|u))\le 
\mathrm{tr} P_{\mathrm{pp}}(L_-(P)) E_{\Sigma_-(P)}(L_-(P))=2 
\end{align*}  
by the min-max principle.  Applying (\ref{Rotation}), one has 
that \[
\mathrm{tr} P_{\mathrm{pp}}(H(P)) E_{\Sigma_-(P)}(H(P))=
\mathrm{tr} P_{\mathrm{pp}}(H(|P|u)) E_{\Sigma_-(P)}(H(|P|u))\le 2.
\]
Thus $H(P)$ has  at most two eigenstates with corresponding eigenvalue 
less than $\Sigma_-(P)$. On the other hand, one already 
knows  that $E(P)< \Sigma_-(P)\le E_1(P)$  for $e<e_*$
by   (\ref{EUDiff}),  
(\ref{LGap}) and  (\ref{MinMaxResult}). Therefore $E(P)$ is at most 
doubly degenerate. $\Box$

\appendix 
 
\section{ Second quantization and basic inequalities} \label{SecondQ}
Let $\h$ be a complex Hilbert space.
The Fock space   over $\h$ is defined by
\[
\Fock(\h)=\Sumoplus\h^{\otimes_{\mathrm{s}}n},
\]
where $\h^{\otimes_{\mathrm{s}}n}$ means the $n$-fold symmetric 
tensor
product of $\h$ with the convention 
$\h^{\otimes_{\mathrm{s}}0}=\BbbC$.
The vector $\Omega=1\oplus0\oplus\cdots\in\Fock(\h)$ is called the Fock
vacuum.

We denote by $a(f)$ the annihilation operator on $\Fock(\h)$ with a  test
vector $f\in \h$, its adjoint  $a(f)^*$, called the creation operator, is defined by 
\[
a(f)^*\vphi=\Sumoplus \sqrt{n+1}f\otimes_{\mathrm{s}}\vphi^{(n)}
\]
for a suitable $\vphi=\sum^{\oplus}_{n\ge 0}\vphi^{(n)}\in \Fock(\h)$.
 By definition, $a(f)$ is 
densely
defined, closed, and antilinear in $f$.  We frequently
write $a(f)^{\#}$ to denote either $a(f)$ or $a(f)^*$.  Creation and
annihilation operators satisfy the canonical
commutation relations
\[
[a(f),a(g)^*]=\la f,g\ra_{\h}\one,\ \ \ 
[a(f),a(g)]=0=[a(f)^*,a(g)^*]
\]
on a suitable subspace of $\Fock(\h)$,
where
$\one$ denotes the identity operator.
We  introduce a particular
subspace of $\Fock(\h)$ which will be used frequently. Let  $\mathfrak{s}$ be a subspace of $\h$.
We define
\[
\Ffin(\mathfrak{s})=\mathrm{Lin}\{a(f_1)^*\dots a(f_n)^*\Omega,\
\Omega\, |\, f_1,\dots,f_n\in\mathfrak{s},\ n\in\BbbN\},
\]
where $\mathrm{Lin}\{\cdots\}$ means the linear span  of the  set
$\{\cdots\}$.
If $\mathfrak{s}$ is dense in $\h$, so is $\Ffin(\mathfrak{s})$ in 
$\Fock(\h)$.

For a densely defined closable operator $c$ on $\h$, 
$\dG(c):\Fock(\h)\to
\Fock(\h)$ is defined by
\begin{align}
\dG(c)\upharpoonright \D(c)^{\otimes_{\mathrm{s}}n}=\sum_{j=1}^{n}
\one\otimes\cdots\otimes\underset{j\,  \mathrm{th}}{c}\otimes 
\cdots\otimes\one\label{dGamma}
\end{align}
and
\[
\dG(c)\Omega=0
\]
where $\D(c)$ means the domain of the
linear operator $c$.
Here in the $j$-th summand $c$ is at the $j$-th entry.
Clearly $\dG(c)$ is closable and we denote its closure by the same
symbol. As a typical   example,  the number operator $N_{\mathrm{f}}$ is 
given
by  $N_{\mathrm{f}}=\dG(\one)$.

In the case where $\h =L^2(\BbbR^3\times \{1,2\})$, 
the annihilation and
creation operator  can be expressed as the operator valued distributions  $a(k, \lambda), a(k, \lambda)^*$ by 
\[
a(f)=\sum_{\lambda=1,2}\int_{\BbbR^3}\mathrm{d}k\, \overline{f}(k, \lambda)a(k,\lambda),\ \
a(f)^*=\sum_{\lambda=1,2}\int_{\BbbR^3}\mathrm{d}k\, f(k, \lambda)a(k, \lambda)^*.
\]
Let $F$ be a measurable function on $\BbbR^3$ and let   the multiplication operator associated with $F$ be denoted 
 by the same symbol:   $(F f)(k, \lambda)=F(k)f(k, \lambda)$  for $f\in L^2(\BbbR^3\times\{1,2\})$.
Then one can formally express $\dG(F)$ as 
\[
\dG(F)=\sum_{\lambda=1,2}\int_{\BbbR^3}\mathrm{d}k\, F(k) a(k, \lambda)^*a(k, \lambda).
\]
For $F(k)=\omega(k)$, one has the expression (\ref{HfDef}) of $\Hf=\dG(\omega)$.

\begin{lemm}\label{FundamentalInq} 
One has the following. 
\begin{itemize} 
\item[{\rm (i)}]$\displaystyle \|a(f)\vphi\|\le 
  \|\omega^{-1/2}f\|\|\Hf^{1/2}\vphi\|$. 
\item[{\rm (ii)}] $\displaystyle \|a(f)^* \vphi\|\le 
  \|(1+\omega^{-1/2})f\| \|(\Hf+\one)^{1/2}\vphi\|$. 
\item[{\rm (iii)}] $\displaystyle a(f)+a(f)^* \le \Hf+\|\omega^{-1/2}f\|^2$. 
\item[{\rm (iv)}] $\displaystyle \|(a(f)+a(f)^*)\vphi\|\le 2 
  \|(1+\omega^{-1/2})f\| \|(\Hf+\one)^{1/2}\vphi\|$. 
\item[{\rm (v)}] $\displaystyle a(f)^{\#_1} a(g)^{\#_2}\le 
  \|(1+\omega^{-1/2})f\|\|(1+\omega^{-1/2})g\|(\Hf+\one)$. 
\end{itemize} 
\end{lemm}

 \section{Invariant domains}
 
 \begin{lemm}\label{InvariantD}
 Let $A$ be  self-adjoint and $H$ be positive and self-adjoint. Assume the following.
 \begin{itemize}
 \item[{\rm (i)}] $(H+\one)^{-1}\D(A)\subseteq \D(A)$.
 \item[{\rm (ii)}] $|\la Hu, Au\ra-\la Au, Hu\ra|\le C \|(H+\one)u\|^2$ for all $u\in \D(A)\cap \D(H)$.
 \item[{\rm (iii)}] $[H, A](H+\one)^{-1}$ can be extended to a bounded operator.
 \end{itemize}
 Then one has $\ex^{\im t A}\D(H)=\D(H)$ for all $t\in \BbbR$.
 \end{lemm}
 {\it Proof.} See \cite[Lemma 2]{GG}. $\Box$

\section{Localization estimate}\label{ProofLocalization} 
In this appendix, we will establish  Lemma \ref{Localization} which is 
essential for the proof of Theorem \ref{Spectrum}. Unfortunately the 
proof  is technically complicated because of the square root 
structure. We repeat  the statement which we want to  
prove. 
\medskip\\ 
{\bf Lemma \ref{Localization} }{\it Choose $e$ as $e<e_*$. For all $\vphi\in 
 \BbbC^2\otimes \Fock_{\mathrm{fin}}\otimes \Fock_{\mathrm{fin}}$, one 
 obtains 
\begin{align*} 
\big|\big\la \vphi, (H(P)-\BGamma(\J)^*H^{\otimes}(P)\BGamma(\J))\vphi\big\ra\big|\le 
 \mathcal{O}(R^0) \|(H(P)+\one)\vphi\|^2,
\end{align*} 
where $\mathcal{O}(R^0)$ is a function of $R$  vanishing as $R\to \infty$. 
}\medskip\\ 
{\it Proof.} 
Let us define a Dirac operator by 
\begin{align*} 
D^{\otimes}(P)&=\alpha\cdot (P-\Pf\otimes\one-\one\otimes \Pf+e\A\otimes 
 \one)+M\beta. 
\end{align*} 
We also introduce 
\begin{align*} 
\hat{D}(P)=D(P)-M\beta,\ \ \  
\hat{D}^{\otimes}(P)=D^{\otimes}(P)-M\beta. 
\end{align*} 
 
 Using the formula (\ref{Fundamental}), 
one has  
\begin{align*} 
|D(P)|&=\frac{1}{\pi}\int_{M^2}^{\infty}\mathrm{d}t\, \frac{\hat{D}(P)^2+M^2} 
{\sqrt{t-M^2}(t+\hat{D}(P)^2)},\\ 
|D^{\otimes}(P)|&=\frac{1}{\pi}\int_{M^2}^{\infty}\mathrm{d}t\, \frac{\hat{D}^{\otimes}(P)^2+M^2} 
{\sqrt{t-M^2}(t+\hat{D}^{\otimes}(P)^2)}. 
\end{align*} 
Hence 
\begin{align*} 
&|D(P)|-\BGamma(\J)^* |D^{\otimes}(P)|\BGamma(\J)\\ 
&=\frac{1}{\pi}\int_{M^2}^{\infty}\mathrm{d}t\, \sqrt{t-M^2}(t+\hat{D}(P)^2)^{-1} 
\Big\{\hat{D}(P)G(P)+G(P)\hat{D}(P)-G(P)^2\Big\}\\ 
&\ \ \ \ \ \ \times(t+\Dirac^2)^{-1}, 
\end{align*} 
where  
\[ 
 G(P)=\hat{D}(P)-\Dirac 
\] 
with 
$\Dirac=\BGamma(\J)^*\hat{ D}^{\otimes}(P)\BGamma(\J)$. 
Remark the following fact 
\[ 
 \|G(P)(\Nf+\one)^{-1}\|\le \mathcal{O}(R^0), 
\] 
see, e.g., \cite{LLG,  LMS}. (It should be noted that the positive photon 
 mass is crucial  here.)  By Lemma \ref{UsefulEst} below, we estimate as  
\begin{align*} 
&\Big\|(\Nf+\one)^{-1}(t+\hat{D}(P)^2)^{-1} 
\Big\{\hat{D}(P)G(P)+G(P)\hat{D}(P)-G(P)^2\Big\}\\ 
&\ \ \ \ \times(t+\Dirac^2)^{-1}(\Nf+\one)^{-1}\Big\|\\ 
&\le 2\|(\Nf+\one)^{-1} (t+\hat{D}(P)^2)^{-1}(\Nf+\one)\| 
 \|(\Nf+\one)^{-1}G(P)\| \|(\Nf+\one)^{-1}\hat{D}(P)\|\\ 
&\ \ \ \ \times \|(\Nf+\one)(t+\Dirac^2)^{-1}(\Nf+\one)^{-1}\|\\ 
&+\|(\Nf+\one)^{-1} (t+\hat{D}(P)^2)^{-1}(\Nf+\one)\| 
 \|(\Nf+\one)^{-1}G(P)\|^2 \\ 
&\ \ \ \ \times \|(\Nf+\one)(t+\Dirac^2)^{-1}(\Nf+\one)^{-1}\|\\ 
&\le \mathcal{O}(R^0)(t^{-2}+t^{-3/2}+t^{-1})^2 
\end{align*}   
which implies 
\begin{align} 
\Big\|(\Nf+\one)^{-1}\big(|D(P)|-\BGamma(\J)^* 
 |D^{\otimes}(P)|\BGamma(\J)\big)(\Nf+\one)^{-1}\Big\| 
\le \mathcal{O}(R^0). \label{NumberEstA} 
\end{align}  
We also  note the fact  
\begin{align} 
 \|(\Nf+\one)^{-1} (\Hf-\BGamma(\J)^* 
\big(\Hf\otimes\one+\one\otimes\Hf)\BGamma(\J)\big)(\Nf+\one)^{-1}\|\le 
 \mathcal{O}(R^0)\label{NumberEstB} 
\end{align} 
which is proven in \cite{LLG}.  
 Collecting (\ref{NumberEstA}) and 
(\ref{NumberEstB}), one sees that $|\la \vphi, 
(H(P)-\BGamma(\J)^*H^{\otimes}(P)\BGamma(\J))\vphi\ra|\le 
\mathcal{O}(R^0)\|(\Nf+\one)\vphi\|^2$ holds.

Finally one has to show $\|(\Nf+\one)\vphi\|\le C 
\|(H(P)+\one)\vphi\|$.  
The positive photon mass implies  $\|(\Nf+\one)\vphi\|\le 
C\|(\Hf+\one)\vphi\|$. Applying Lemma 
\ref{AppB1} yields the desired results 
$\|(\Nf+\one)\vphi\|\le C 
\|(H(P)+\one)\vphi\|$.  
 $\Box$

\begin{lemm}\label{UsefulEst} 
For all $t>0$, one has the following. 
\begin{itemize} 
\item[{\rm (i)}] $\displaystyle 
                 \|(\Nf+\one)(t+\hat{D}(P)^2)^{-1}(\Nf+\one)^{-1}\|\le C(t^{-1}+t^{-3/2}+t^{-2}). $  
\item[{\rm (ii)}] $\displaystyle \|(\Nf+\one)(t+\Dirac^2)^{-1}(\Nf+\one)^{-1}\|\le C(t^{-1}+t^{-3/2}+t^{-2}). $ 
\end{itemize} 
\end{lemm}  
{\it Proof.} (i)  The essential idea is taken from \cite{LL3}. 
First we will show  that \\$\ex^{\im t \hat{D}(P)}\D(\Nf)=\D(\Nf)$.
It suffices to check the conditions (i), (ii) and (iii) in Lemma \ref{InvariantD}.
Noting $[\Nf, \hat{D}(P)]=-\alpha\cdot (a(F_0)-a(F_0)^*)$ on $\BbbC^4\otimes \Ffin$, we can check all conditions in Lemma
\ref{InvariantD} by Lemma \ref{FundamentalInq}.

Using the formula 
\begin{align*} 
(\hat{D}(P)^2+t)^{-1}=\int_{\BbbR}\mathrm{d}s\, g_t(s)\, \ex^{-\im s \hat{D}(P)} 
\end{align*} 
with $g_t(s)=\sqrt{\pi/2t}\, \ex^{-\sqrt{t}|s|}$, we have 
\begin{align} 
\|(\Nf+\one)(\hat{D}(P)^2+t)^{-1}\vphi\|\le \int_{\BbbR}\mathrm{d}s\, 
 g_t(s)\|(\Nf+\one)\ex^{-\im s\hat{D}(P)}\vphi\|\label{IntI} 
\end{align} 
for each normalized $\vphi\in \D(\Nf)$. (We already know that $\ex^{\im s\hat{D}(P)}\vphi\in \D(\Nf).$) 
Set  $I_1(s)=\|(\Nf+\one)\ex^{-\im s\hat{D}(P)}\vphi\|$ and 
$I_{1/2}(s)=\|(\Nf+\one)^{1/2}\ex^{-\im s \hat{D}(P)}\vphi\|$.  
Then one has  
\begin{align*} 
\frac{\mathrm{d}}{\mathrm{d}s} I_1(s)^2&=\big\la \ex^{-\im s 
 \hat{D}(P)}\vphi, \im [\hat{D}(P), (\Nf+\one)^2]\ex^{-\im s 
 \hat{D}(P)}\vphi\big\ra\\ 
&=\la \ex^{-\im s\hat{D}(P)}\vphi, \big(e\alpha\cdot E 
 (\Nf+\one)+(\Nf+\one)e\alpha\cdot E\big)\ex^{\im s \hat{D}(P)}\vphi\big\ra, 
\end{align*}  
where $E=\im a(F_0)-\im a(F_0)^*$. Accrodingly using the standard 
estimate $\||E|\vphi\|\le C\|(\Nf+\one)^{1/2}\vphi\|$, one has  
\begin{align} 
\frac{\mathrm{d}}{\mathrm{d}s} I_1(s)^2\le C I_{1/2}(s)I_1(s).\label{IEst} 
\end{align}  
Next we will estimate $I_{1/2}(s)$. Observe that   
\begin{align*} 
\frac{\mathrm{d}}{\mathrm{d}s} I_{1/2}(s)^2&=\big\la \ex^{-\im s 
 \hat{D}(P)}\vphi, \im [\hat{D}(P), \Nf]\ex^{-\im s 
 \hat{D}(P)}\vphi\big\ra\\ 
&\le e\||E|\ex^{-\im s\hat{D}(P)}\vphi\|\\ 
&\le  C \|(\Nf+\one)^{1/2}\ex^{-\im s \hat{D}(P)}\vphi\|\\ 
&= C I_{1/2}(s). 
\end{align*} 
Solving this ineqaulity,  we get $I_{1/2}(s)\le C|s| 
+I_{1/2}(0)$. Inserting this result into (\ref{IEst}), one has 
\begin{align*} 
I_1(s)\le I_1(0)+Cs^2 +C|s| I_{1/2}(0). 
\end{align*}  
Combining this with (\ref{IntI}), we finally obtain the assertion (i) in the 
lemma.  
 
Noting  the fact 
$\BGamma(\J)\Nf\BGamma(\J)^*=\Nf\otimes\one+\one\otimes \Nf$, 
one can apply the similar arguments in the proof of (i) to show (ii). 
$\Box$

\section{Auxiliary estimates} 
 
In this appendix, we always choose $e$ as $e<e_*$.

\begin{lemm}\label{AppB02} 
For all $P\in \BbbR^3$ and $e\ge 0$, one has  
\begin{align*} 
\||D(P)|(\Hf+\one)^{-1}\|\le |P| +3+3e \|\omega^{-1/2} F_0\|. 
\end{align*}  
\end{lemm} 
{\it Proof.} Noting Lemma \ref{FundamentalInq} and  the fundamental fact 
$ 
\||P_{\mathrm{f}, 
i}|(\Hf+\one)^{-1}\|\le 1 
$ 
 for $i=1,2,3$, one observes that  
\begin{align*} 
&\||D(P)|(\Hf+\one)^{-1}\|\\ 
&=\|D(P)(\Hf+\one)^{-1}\|\\ 
&\le  |P|+\sum_{i=1,2,3}\||P_{\mathrm{f}, 
 i}|(\Hf+\one)^{-1}\|+e\sum_{i=1,2,3}\|\A_i(\Hf+\one)^{-1}\|\\ 
&\le |P|+3+3e \|\omega^{-1/2}F_0\|. 
\end{align*}  
This proves the assertion. $\Box$

\begin{lemm}\label{AppB1} 
For each $n\in\BbbN$, one obtains 
\begin{align*} 
 \|\Hf^{n/2}(H(P)+\one)^{-n/2}\|&\le \C,\\ 
\||D(P)|^{n/2}(H(P)+\one)^{-n/2}\|&\le \C, 
\end{align*} 
where $\C$ is independent of $P$. 
\end{lemm} 
{\it Proof.} In the similar way in the proof of \cite{FGS2}, one can show that  
both  
\begin{align} 
\Hf^{n/2}(H+\one)^{-n/2},\ \  |D|^{n/2}(H+\one)^{-n/2} 
\end{align} 
 are bounded. Thus we  conclude the assertion by the fact that $ 
(H(P)+\one)^{n/2}\vphi$ is strongly continuous in $P$ for $\vphi\in 
\BbbC^4\otimes \Ffin$. $\Box$

\begin{lemm}\label{AppB2} 
For each $n\in\BbbN$, we obtain 
\begin{align*} 
\|(\Hf+\one)^{n}(t+\hat{D}(P)^2)^{-1}(H(Q)+\one)^{-n}\|\le \C 
 (t^{-1}+t^{-3/2}+\cdots + t^{-n-1}) 
\end{align*} 
for every $P, Q\in \BbbR^3$, where $\C$ is independent of $P$ and $Q$. 
\end{lemm} 
{\it Sketch of proof.}
By Lemma \ref{FundamentalInq},  one can see that $\ex^{\im t \hat{D}(P)}\D(\Hf^n)=\D(\Hf^n)$.
 Let us write 
\[ 
 K_{m/2}(s) =\|(\Hf+\one)^{m/2}\ex^{-\im s \hat{D}(P)}\vphi\| 
\] 
for a normalized $\vphi \in \D(\Hf^n)$ with $m\le 2n$. In the case where  $m=1$, one has  
\begin{align*} 
\frac{\mathrm{d}}{\mathrm{d}s} K_{1/2}(s)^2&=\la \ex^{-\im s 
 \hat{D}(P)}\vphi, \im [\hat{D}(P), \Hf]\ex^{-\im s 
 \hat{D}(P)}\vphi\ra\\ 
&=\la \ex^{-\im s \hat{D}(P)}\vphi, \im \alpha\cdot (a(\omega 
 F)-a(\omega F)^*)\ex^{-\im s \hat{D}(P)}\vphi\ra\\ 
&\le C K_{1/2}(s) 
\end{align*} 
by the Schwarz inequality. Thus $K_{1/2}(s)\le K_{1/2}(0)+C|s|$ holds. 
In the case where  $m=2$, one has, by the similar arguments in the above,  
\begin{align*} 
\frac{\mathrm{d}}{\mathrm{d}s} K_1(s)^2&\le  C K_{1/2}(s) K_1(s)\\ 
&\le (K_{1/2}(0)+C|s|)K_1(s)  
\end{align*}  
which implies $K_1(s)\le K_1(0)+C(K_{1/2}(0)|s|+s^2)$. Repeating this 
procedure, one can arrive at  
\[ 
 K_{m/2}(s)\le K_{m/2}(0)+C (K_{(m-1)/2}(0)|s|+\cdots+ K_{1/2}(0)|s|^{m-1}+|s|^m). 
\] 
Therefore using the formula  
\begin{align*} 
\|(\Hf+\one)^{m/2}(t+\hat{D}(P)^2)^{-1}\vphi\|\le 
 \int_{\BbbR}\mathrm{d}s\, g_t(s)K_{m/2}(s) 
\end{align*} 
with $g_t(s)=\sqrt{\pi/2t}\ex^{-\sqrt{t}|s|}$, one has , by putting $m=2n$, 
\begin{align}
 \|(\Hf+\one)^{n}(t+\hat{D}(P)^2)^{-1}(\Hf+\one)^{-n}\|\le C 
 (t^{-1}+t^{-3/2}+\cdots +t^{-n-1}).\label{HfDHf} 
\end{align}
 Finally using Lemma \ref{AppB1}, one  concludes the desired assertion 
 in the lemma. $\Box$ 
 
\begin{lemm}\label{AppB3}One has  
\[ 
 \|[\Hf, |D(P)|](H(Q)+\one)^{-2}\|\le \C (1+|P|) 
\] 
for all $P, Q\in \BbbR^3$. 
\end{lemm} 
{\it Proof.} By Lemma \ref{AppB2} and the following standard formula 
\begin{align*} 
 [\Hf, |D(P)|]=\frac{1}{\pi}\int_{M^2}^{\infty}\mathrm{d}s\, 
 \sqrt{s-M^2}(s+\hat{D}(P)^2)^{-1} 
[\Hf, \hat{D}(P)^2](s+\hat{D}(P)^2)^{-1}, 
\end{align*} 
one computes 
\begin{align*} 
&\|[\Hf, |D(P)|](\Hf+\one)^{-2}\|\\ 
&\le  \frac{1}{\pi}\int_{M^2}^{\infty}\mathrm{d}s\, 
 \sqrt{s-M^2}\|(s+\hat{D}(P)^2)^{-1}\| 
\|[\Hf, \hat{D}(P)^2](\Hf+\one)^{-2}\|\\ 
&\ \ \ \ \times\|(\Hf+\one)^2 (s+\hat{D}(P)^2)^{-1}(H(Q)+\one)^{-2}\|\\ 
&\le \frac{C}{\pi}\int_{M^2}^{\infty}\mathrm{d}s\, 
 \sqrt{s-M^2}s^{-1}(1+|P|)(s^{-1}+s^{-3/2}+\cdots+s^{-3})\\ 
&= \C (1+|P|).  
\end{align*}  
This completes the proof. $\Box$ 
 
\begin{lemm}\label{AppB4} 
For all $P, Q\in\BbbR^3$, one has  
\begin{align*} 
&\|H(P)^2(H(Q)+\one)^{-2}\|\\ 
&\le  (C+|P-Q|)^2+C (1+|P|+|Q|+|P||Q|)(C+|P-Q|^2) 
\end{align*}  
\end{lemm} 
{\it Proof.} First we will show that  
\[ 
 \|H(P)(H(Q)+\one)^{-1}\|\le C+|P-Q|. 
\] 
To this end, observe that  
\begin{align*} 
&\|H(P)(H(Q)+\one)^{-1}\|\\ 
&\le  \||D(P)|(H(Q)+\one 
 )^{-1}\|+\|\Hf(H(Q)+\one)^{-1}\|\\ 
&\le \|D(Q)(H(Q)+\one)^{-1}\|+\|\alpha\cdot 
 (P-Q)(H(Q)+\one)^{-1}\|+\|\Hf(H(Q)+\one)^{-1}\|\\ 
&\le  C+|P-Q| 
\end{align*} 
by Lemma \ref{AppB1}. 
Write  
\begin{align*} 
&H(P)^2(H(Q)+\one)^{-2}\\ 
&= \{ H(P)(H(Q)+\one)^{-1}\}^2+H(P)[H(P), 
 (H(Q)+\one)^{-1}](H(Q)+\one)^{-1}\\ 
&= \{ H(P)(H(Q)+\one)^{-1}\}^2+H(P)(H(Q)+\one)^{-1}[H(Q),H(P)](H(Q)+\one)^{-2}. 
\end{align*} 
Hence  
\begin{align*} 
&\|H(P)^2(H(Q)+\one)^{-2}\|\\ 
& \le  (C+|P-Q|)^2+(C+|P-Q|)\|[H(Q), H(P)](H(Q)+\one)^{-2}\|. 
\end{align*} 
Accordingly what we have to show next is to estimate  
the operator norm \[ 
                   \|[H(Q), H(P)](H(P)+\one)^{-2}\|. 
\] 
Observe that  
\begin{align} 
&[H(Q), H(P)](H(Q)+\one)^{-2}\nonumber\\ 
=&[|D(P)|, |D(Q)|](H(Q)+\one)^{-2}\label{HPQ1}\\ 
&+[\Hf, |D(P)|](H(Q)+\one)^{-2}\label{HPQ2}\\ 
&+[|D(P)|, \Hf](H(Q)+\one)^{-2}\label{HPQ3}. 
\end{align}  
Norm of (\ref{HPQ2}) and (\ref{HPQ3}) can be  estimated by Lemma 
\ref{AppB3}. As to (\ref{HPQ1}), note that  
\begin{align*} 
&\||D(P)||D(Q)|\vphi\|\\ 
&\le  \||D(0)||D(Q)|\vphi\|+|P|\||D(Q)|\vphi\|\\ 
&\le  C(\|(\Hf+\one)|D(Q)|\vphi\|+|P|\||D(Q)|\vphi\|)\\ 
&\le C \Big(\||D(Q)|(\Hf+\one)\vphi\|+\|[\Hf, 
 |D(Q)|]\vphi\|+(1+|Q|)|P|\|(\Hf+\one)\vphi\|\Big)\\ 
&\le C 
 \Big((1+|Q|)\|(\Hf+\one)^2\vphi\|+\|(H(Q)+\one)^2\vphi\|+(1+|Q|)|P|\|(\Hf+\one)\vphi\|\Big)\\ 
&\le  C (1+|P|+|Q|+|P||Q|)\|(H(Q)+\one)^2\vphi\|. 
\end{align*} 
In the above  we have used Lemma \ref{AppB3} from the  line three  to the next, and 
from  the line four  to the final line, we have used Lemma \ref{AppB02}.  
Collecting the results, one obtains the assertion in the lemma. $\Box$ 
 
\begin{lemm}\label{AppB5} 
Let  
\begin{align*} 
\s(P)=|D(P-k)|a(k, \lambda)-a(k, \lambda)|D(P)|. 
\end{align*}  
Then one has  
\begin{align*} 
\|\s(P)^*(H(P-k)+\one)^{-1}\|\le C (1+|k|)|F_0(k, \lambda)|, 
\end{align*}  
where $C$ is independent of $k$ and $P$. 
\end{lemm}  
{\it Proof.} We will show that $\|(H(P-k)+\one)^{-1} S_{k,
 \lambda}\|\le C(1+|k|)|F_0(k, \lambda)|$.
 Let  
\begin{align*} 
\s= \ex^{\im k\cdot x} \ex^{\im x\cdot \Pf} [|D|, a(k, 
 \lambda)]\ex^{-\im x\cdot \Pf}. 
\end{align*} 
Then one has  
\[ 
 \mathcal{F}_x \s \mathcal{F}^*_x =\int^{\oplus}_{\BbbR^3}\s(P)\, 
 \mathrm{d}P 
\] 
for all $(k, \lambda)\in \BbbR^3\times \{1,2\}$. Hence it suffices to 
show that  
\begin{align} 
\|(H_k+\one)^{-1}[|D-\alpha\cdot k|, a(k, \lambda)]\|\le C (1+|k|)|F_0(k, \lambda)|\label{AppB5-1} 
\end{align} 
where $H_k=|D-\alpha\cdot k|+\Hf$, because  
\begin{align*} 
&\mathcal{F}^*_x\int^{\oplus}_{\BbbR^3}(H(P-k)+\one)^{-1}\s(P)\, 
 \mathrm{d}P\, \mathcal{F}_x\\ 
&= 
 \ex^{\im x\cdot \Pf}(H_k+\one)^{-1}\ex^{-\im x\cdot \Pf}\s\\ 
&=  \ex^{\im x\cdot \Pf}(H_k+\one)^{-1}\ex^{-\im 
 x\cdot \Pf}\ex^{\im k\cdot x }\ex^{\im x\cdot \Pf}[|D|, a(k, \lambda)]\ex^{-\im x\cdot 
 \Pf}\\ 
&= \ex^{\im x\cdot \Pf}(H_k+\one)^{-1}[|D-\alpha\cdot k|, a(k, \lambda)]\ex^{\im k\cdot
 x}\ex^{-\im x\cdot \Pf}. 
\end{align*}  
To this end, we remark that, with $\hat{D}=D-M\beta$,  
\begin{align} 
&[|D-\alpha\cdot 
 k|, a(k, \lambda)]\nonumber\\ 
&=\frac{1}{\pi}\int_{M^2}^{\infty}\mathrm{d}s\, 
 \sqrt{s-M^2}\big(s+(\hat{D}-\alpha\cdot k)^2\big)^{-1}[ 
 (\hat{D}-\alpha\cdot k)^2, a(k, \lambda)]\nonumber\\ 
&\ \ \ \ \times\big(s+(\hat{D}-\alpha\cdot 
 k)^2\big)^{-1}\nonumber\\ 
&=-\frac{1}{\pi}\int_{M^2}^{\infty}\mathrm{d}s\, 
 \sqrt{s-M^2}\big(s+(\hat{D}-\alpha\cdot 
 k)^2\big)^{-1}\Big\{(\hat{D}-\alpha\cdot k)e\alpha\cdot F_{x}(k, 
 \lambda)\nonumber\\ 
&\ \ \ \ +e\alpha\cdot F_{x}(k, \lambda)(\hat{D}-\alpha\cdot 
 k)\Big\}\big(s+(\hat{D}-\alpha\cdot k)^2\big)^{-1}.\label{AppB5-2} 
\end{align} 
Assume, for a while,  that 
\begin{align} 
\|(H_k+\one)\big(s+(\hat{D}-\alpha\cdot 
 k)^2\big)^{-1}(H_k+\one)^{-1}\|\le C(s^{-1}+s^{-3/2}+s^{-2}).\label{AppB5-3} 
\end{align} 
We note  the following two estimates:  
\begin{align} 
&\|(H_k+\one)^{-1}(\hat{D}-\alpha\cdot 
 k)\alpha\cdot F_{x}(k, \lambda)\|\nonumber\\ 
&=\|(H+\one)^{-1}\hat{D}\alpha\cdot F_{0}(k, \lambda)\|\nonumber\\ 
&\le  C|F_0(k, \lambda)|\label{AppB5-4} 
\end{align}  
and 
\begin{align} 
&\|(H_k+\one)^{-1}\alpha\cdot F_{x}(k, 
 \lambda)(\hat{D}-\alpha\cdot k)\|\nonumber\\ 
&=\|(H+\one)^{-1}\alpha\cdot F_0(k, \lambda)(\hat{D}+\alpha\cdot k)\|\nonumber\\ 
&\le |F_0(k, \lambda)|\|(H+\one)^{-1}\hat{D}\|+|k||F_0(k, 
 \lambda)|\|(H+\one)^{-1}\|\nonumber\\ 
&\le  C(1+|k|)|F_0(k, \lambda)|.\label{AppB5-5} 
\end{align} 
because $\hat{D}(H+\one)^{-1}$ is bounded and  $\ex^{-\im k\cdot 
x}H_k=H\ex^{-\im k\cdot x}, \ex^{-\im k \cdot x}(\hat{D}-\alpha\cdot 
k)=\hat{D}\ex^{-\im k\cdot x}$. 
Collecting (\ref{AppB5-2}), (\ref{AppB5-3}), (\ref{AppB5-4}) and 
(\ref{AppB5-5}), one has  
\begin{align*} 
&\|[|D-\alpha\cdot k|, a(k, \lambda)](H_k+\one)^{-1}\|\\ 
&\le  \frac{1}{\pi}\int_{M^2}^{\infty}\mathrm{d}s\, 
 \sqrt{s-M^2}s^{-1}C(1+|k|)|F_0(k, \lambda)| (s^{-1}+s^{-3/2}+s^{-2})\\ 
&=  C(1+|k|)|F_0(k, \lambda)|. 
\end{align*} 
This is what we want to show. 
 
Finally we will prove (\ref{AppB5-3}). Basic strategy is similar to the 
proof of Lemma \ref{AppB2}. Let $J_{n/2}(s)=\|(H_k+\one)^{n/2}\ex^{-\im 
s (\hat{D}-\alpha\cdot k)}\vphi\|,\ n=1,2$ for  $\vphi\in \D(H_k)$ with 
$\|\vphi\|=1$. Then, since 
$[\hat{D}-\alpha\cdot k, H_k]=[\hat{D}, \Hf]=\alpha\cdot (a(\omega 
F_x)-a(\omega F_x)^*)$, 
one can easily modify the proof of Lemma \ref{AppB2} to conclude that  
\begin{align*} 
J_1(s)\le J_1(0)+C(J_{1/2}(0)|s|+s^2),\ \ J_{1/2}(s)\le J_{1/2}(0)+C|s|. 
\end{align*} 
Thus, using the formula 
\[ 
 (t+\hat{D}^2)^{-1}=\int_{\BbbR}\mathrm{d}s\, g_t(s)\ex^{-\im s \hat{D}} 
\] 
with $g_t(s)=\sqrt{\pi/2t}\,  \ex^{-\sqrt{t}|s|}$ and modifying the 
proof of Lemma \ref{AppB2}, one can arrive at (\ref{AppB5-3}). $\Box$ 
 
 \section{Operator monotonicity of the square root}
 
 \begin{lemm}\label{OpMono}  {\rm (Operator monotonicity of the square root: Unbounded version)}
Let $S$ and  $T$ be two positive self-adjoint operators (not necessarily bounded) with $\D(S^{1/2})\supseteq \D(T^{1/2})$.
Assume that $S\le T$. Then one has $\D(S^{1/4})\supseteq \D(T^{1/4})$ and $\sqrt{S}\le \sqrt{T}$.
\end{lemm}
{\it Proof.} Set $E_n=E_S([0, n])$ where $E_S(\cdot)$ is the spectral measure of $S$. Define $S_n=E_n^{1/2}SE_n^{1/2}$.
Then one has $S\ge S_n\ge 0$ for all $n\in \BbbN$.  Thus  $S_n\le T$ holds for all $n\in\BbbN$. Now one has 
\begin{align}
(\vepsilon T+\one)^{-1}S_n(\vepsilon T+\one)^{-1}\le (\vepsilon T+\one)^{-1} T (\vepsilon T+\one )^{-1} \label{MonoInq}
\end{align}
for every $\vepsilon>0$.
 Since both sides  of (\ref{MonoInq}) are positive and bounded, one can  apply the operator monotonicity
 of the square root for {\it bounded} positive operators \cite{Pedersen} and obtain
 \begin{align*}
 \sqrt{(\vepsilon T+\one)^{-1} S_n (\vepsilon T+\one)^{-1}} \le \sqrt{(\vepsilon T+\one)^{-1} T (\vepsilon T+\one)^{-1}}
 \end{align*} 
 for all $\vepsilon >0$.
Taking $\vepsilon \downarrow 0$ first, we have $\sqrt{S_n}\le \sqrt{T}$ for each $n\in \BbbN$. 
It follows  that, for $f\in \D(T^{1/4})$, one has 
\begin{align*}
\int_0^n\lambda^{1/2}  \mathrm{d}  \|E_S(\lambda) f\|^2   \le   \langle f,\sqrt{T} f\rangle
\end{align*}
by the spectral theorem.
Now taking $n\to \infty$, 
we conclude that $f\in \D(S^{1/4})$ and  $\langle  f, \sqrt{S}f \rangle  \le \langle  f,  \sqrt{T}f \rangle  $ by the monotone convergence theorem.  $\Box$

\end{document}